\documentclass[showpacs, twocolumn]{revtex4-1}
\usepackage{graphicx}
\usepackage{amsmath}
\usepackage{appendix}
\usepackage{xcolor}
\usepackage{colortbl}
\usepackage{lineno}

\begin{document}

\title{ Self-gravitating Lee-Huang-Yang fluids as dark matter halos}

\author{Abdel\^{a}ali Boudjem\^{a}a }
\affiliation{Department of Physics, Faculty of Exact Sciences and Informatics.\\
Laboratory of Mechanics and Energy, Hassiba Benbouali University of Chlef, P.O. Box 78, 02000, Ouled-Fares, Chlef, Algeria.}
\email {a.boudjemaa@univ-chlef.dz}

\date{\today}


\begin{abstract}

We  propose a model of dark matter  halos  based on a self-gravitating Bose fluid whose effective equation of  state is dominated by Lee-Huang-Yang quantum fluctuations.
The model is governed by coupled generalized Gross-Pitaevskii-Poisson equations, which incorporate quantum pressure, Lee-Huang-Yang-induced repulsion, and gravitational attraction. 
By retaining the full quantum pressure, we examine the halo structure from its central region to its asymptotic regime. 
We first derive the Jeans dispersion relation and analyze how the  Lee-Huang-Yang fluctuations alter the gravitational instability criterion and the associated characteristic length and mass scales.  
We then obtain analytical expressions for the central and large-radius behaviors of the equilibrium solutions.  
To verify the validity and the performance of  the model, we  fit the  model-predicted  curves with a sample of 20 galaxies from the SPARC database and compare the results with those obtained from  Navarro-Frenk-White, standard Bose-Einstein condensate, and quantum-droplet dark matter models. We find that the pure-LHY model reproduces a broad range of observed galactic kinematics, yielding reduced $\chi^2$ values  and a lower sample  average  compared to the aforementioned models. 
These results indicate that the combined effects of  quantum pressure,  LHY quantum fluctuations and the self-gravity  provide additional flexibility in describing  both the central and extended outer regions of galactic halos. 
We also discuss the implications  of the inferred halo parameters and emphasize their dependence on the assumed effective equation of state and its microscopic interpretation.
The pure-LHY  model constitutes an ideal platform for exploring  the role of quantum-fluctuation-induced interaction in the formation and structure of dark matter halos. 
 
\end{abstract}

\maketitle

\section{Introduction} \label{Intro}

The existence of unseen gravitating matter was first inferred by Zwicky from the dynamics of the Coma cluster \cite{Zwicky1933}.
Subsequent observations of flat rotation curves in spiral galaxies \cite{Rubin1970,Bosma1981}  provided compelling evidence for the 
presence of dark matter (DM) on galactic scales, indicating that its gravitational contribution extends well beyond the visible regions of  galaxies.
Rotation-curve studies have since  become one of the principal tools for investigating the distribution of  DM in galaxies \cite{Sofue2001} (for a historical account see, e.g. Ref.~\cite{Bertone2018}). 
Despite this long history,  the nature of DM remains one of the most mysterious problems in  modern cosmology and astrophysics.  
It is widely regarded as a fundamental component of the standard cosmological model, with existence inferred from a broad range of astrophysical and cosmological observations  (see e.g. \cite{Strig,Zurek,Court,Baer,Aram,Arb,Mat}).
Evidence from galactic rotation curves, gravitational lensing, galaxy clusters, and the formation of large-scale cosmic structures indicates that a substantial fraction of the matter content of the Universe is nonluminous and interacts predominantly through gravity
\cite{Rubin1980,Clowe2006,Bertone2005,Planck2020}.

From the particle-particle perspective, although weakly interacting massive particles (WIMPs) remain well-motivated candidates, the absence of conclusive signals in  direct- and indirect-detection experiments, as well as collider searches, has increasingly constrained their simplest realizations and motivated the exploration of alternative DM scenarios  \cite{Arcadi2018,Roszkowski2018}.
The standard $\Lambda$ cold DM ($\Lambda$-CDM) cosmological model provides a successful framework for describing the evolution of cosmic structures and the observed large-scale properties of the Universe.  
Nevertheless,  the internal structure of DM halos on galactic and subgalactic scales continues to  motivate investigations of alternative DM models \cite{Peed,Pad,Cop,Nor,Wen,Noj}.
In particular  discrepancies between some observed galactic density profiles and the central density cusps predicted by collisionless CDM simulations have stimulated extensive discussion of the core-cusp problem and the diversity of galactic rotation curves 
\cite{Navarro1996,deBlok2010,Bertone2005}.  Although baryonic feedback and other astrophysical processes may influence halo structure, the extent to which these processes explain the full range of observed galactic properties remains an active subject of research. 

Galactic rotation curves provide an important observational probe of the distribution of matter in galaxies. For a spherically symmetric halo, the circular velocity is related to the enclosed mass through $v_{\mathrm{tg}}(r)=\sqrt{GM(r)/r}$, where $G$ is the Newton's gravitational constant and $M(r)$  is the mass enclosed with radius $r$. The approximately flat rotation curves observed in many spiral  galaxies imply that the enclosed gravitational mass continues to increase with radius over the measured  region, beyond  what would be expected  from the luminous component alone  \cite{Rubin1980,Persic1996}.  The Navarro-Frenk-White (NFW) density profile motivated by cosmological simulations of collisionless  $\Lambda$-CDM, has become a standard reference model for describing  DM halos  \cite{Navarro1996}. However, its central cusp and the diversity of its predicted inner rotation curves have prompted the development of alternatives involving self-interaction, wave effects, and quantum pressure.  Therefore, a viable halo should be assessed not only by its ability to reproduce a particular rotation curve, but also by the physical consistency of its density profile, its  asymptotic behavior, and the number of adjustable parameters required to describe the  data.

A prominent framework models  DM  in terms of ultralight bosonic particles whose macroscopic dynamics can be represented by a classical field or an effective quantum fluid.  When the  de-Broglie wavelength of the bosons becomes astrophysically relevant, wave effects can modify the structure of DM halos and suppress the formation of small-scale density fluctuations  \cite{Hu2000,Schive2014, Hui2017}. 
In the  nonrelativistic  regime, the evolution of such systems is frequently described by coupled Schr\"odinger-Poisson or Gross-Pitaevskii-Poisson equations  (GPPE), depending on whether self interactions are included explicitly. 
Quantum gradient contributions can oppose gravitational compression and produce central density profiles that differ from those of classical collisionless matter.  
Bose-Einstein condensed DM (BEC DM) models provide another framework in which quantum statistics and particle interactions influence halo structure and the resulting  rotation curves \cite{Bohmer2007,Chavanis2011}. 
Related approaches consider ultralight scalar fields and axion BEC, exploring the role  of  wave effects in the formation and structure of DM halos \cite{Hu2000,Hui2017,SikivieYang2009,Harko2011}.
The  predictions of these models  depend on the assumed microscopic interaction, the equation of state (EOS), the boundary conditions, and the relative importance of quantum pressure and interactions.

Beyond the mean-field approximation, quantum fluctuations furnish an additional mechanism for developing effective description of interacting Bose fluids. 
The Lee-Huang-Yang (LHY) correction originates from quantum fluctuations  beyond leading-order mean-field  theory and accounts for their contribution to the EOS of a dilute  Bose gas  \cite{LHY}.
In ultracold atomic systems, LHY contributions have attracted considerable attention because they can compete with mean-field interaction and stabilize quantum droplets under appropriate conditions \cite{Petrov2015}. 
This motivates investigating the role of LHY-type nonlinearities in other settings where quantum effects and self-gravity are relevant.  However, an important distinction must be made between the conventional LHY correction to an interacting Bose gas and an effective fluid whose dominant  interaction term has LHY-type density dependence.  The standard  LHY term does not generally replace the leading mean-field contribution on its own. Therefore,  an LHY-dominant  DM  model requires a clear  statement  of the physical assumption under which the mean-field contribution  can be neglected, suppressed, or compensated. In the present work, we investigate a  self-gravitating  Bose-fluid with a pure-LHY effective EOS,  as can occur in ultradilute  binary BEC mixtures near the mean-field balance \cite{Petrov2015,Boudjemaa2026}. Whether such a mixture can realize DM halos is treated as a phenomenological hypothesis to be tested against  rotation-curve data.

It is useful to distinguish the present pure-LHY DM model from our previously proposed quantum-droplet DM (QD2M) framework \cite{Boudjemaa2026}  and the conventional BEC DM model \cite{Bohmer2007,Chavanis2011}. 
In the latter, the condensate is generally  described by the GPPE with two-body  contact interaction, leading to a mean field chemical potential  proportional to the particle density  $(\propto n)$ and  the interaction pressure $\propto n^2$. 
In the Thomas-Fermi (TF) regime, this corresponds to a polytropic equation with index $p=1$  \cite{Bohmer2007,Chavanis2011}. By contrast, the QD2M framework considers a Bose mixture in which an attractive mean-field contribution competes with a repulsive LHY corrections, giving rise to a stabilization of a self-bound droplet. This  exotic state is governed by an extended EOS whose properties depend on the interspecies interaction and on the strength of quantum fluctuations \cite{Boudjemaa2026}.
The present model investigates a different effective regime: a single-component, LHY-dominated fluid with chemical potential  is $\propto n^{3/2}$  and pressure  $\propto n^{5/2}$. Its corresponding polytropic  index is  $p=2/3$, and its halo structure is determined by the interplay between this effective pressure, self-gravity, and the full quantum gradient term retained in generalized GPPE (GGPPE).  
Thus, unlike QD2M, the model  does not explicitly rely  on a balance between attractive mean-field and repulsive LHY terms to form a self-bound droplets, and the standard BEC DM, its leading effective pressure scales as  $n^{5/2}$ rather than $n^{2}$.
This difference changes the relation between density and pressure and consequently modifies the hydrostatic balance between gravitational attraction, interaction pressure, and quantum gradient effects. 
It is therefore important to determine whether the LHY-dominated regime can generate halo configurations with density profiles and circular-velocity distributions comparable with galactic observations. 
Several extensions of GPPE  have been introduced to explain  the internal structures of BEC DM,  including  generalized nonlinear wave equations with dissipation, models combining a solitonic core with an isothermal atmosphere,  
coupled condensate–particle descriptions  and  logarithmic nonlinearity \cite{Chavanis2019,Proukaki6,Daller,HaghaniHarko2026}.        

The cosmological motivation is related to the role of the effective EOS in the gravitational evolution of density perturbations. The pressure and compressibility of a bosonic DM influence its characteristic Jeans scale and can hence affect the scales on which gravitational perturbations grow or are suppressed.  
A pure LHY EOS provides a different density dependence from that of standard contact interaction BEC DM, motivating a dedicated analysis of its linear stability, characteristic length scales, and possible implications for halo formation. 
Nevertheless, changes in the Jeans scale do not by themselves establish agreement with cosmological observations:  a complete  cosmological assessment would require the appropriate perturbation equations, the evolution of the background  density, 
and comparison with constraints from structure formation and other cosmological probes. 
At the galactic  scale, rotation-curve fitting provides a complementary test of whether the resulting equilibrium solutions can describe observed systems.  
The pure-LHY description should therefore be regarded  as a distinct effective halo model,  with its microscopic validity depending on the conditions under which the conventional mean-field contribution can be neglected.

Recent studies have examined LHY effects in QD2M  halos \cite{Boudjemaa2026}  and condensate dark stars \cite{Panotopoulos2026LHY}. 
In contrast,  the present work addresses a different question: whether a fluid dominated by an LHY-type contribution,  with  $(P\propto n^{5/2})$, can support extended galactic halo configurations.
Its purpose is therefore not to assume that the LHY correction necessarily describes DM, but to establish the structural and observational consequences of this particular effective model and identify the conditions under which it may be astrophysical relevant. 
To this end, we self-consistently solve the full GGPPE  which incorporates  the quantum pressure, the effective LHY nonlinearity and gravitational attraction.  It also permits a systematic investigation of the central density  behavior, the asymptotic density  tail, and the condition under which a spatially extended halo can form.  
Furthermore, a linear Jeans analysis \cite{Jean} is used to  examine the stability of homogeneous perturbations and to identify the characteristic scales associated with gravitational collapse in the fluid. 
Such an analysis complements the equilibrium  calculation and helps assess the model's potential cosmological relevance.

The asymptotic behavior of the resulting density profile is particularly important when interpreting galactic rotation curves.  If the outer density decays exponentially,  the total  halo mass can converge to a finite value.  
The corresponding circular velocity then approaches the Keplerian form, \(v (r)\propto r^{-1/2}\), sufficiently far outside the mass distribution.  
An exponentially decaying halo thus cannot sustain an indefinitely flat rotation curve at arbitrarily large radii.  Any approximately flat region must instead occur over a finite radial interval in which the enclosed mass increases sufficiently rapidly with radius. 
This distinction is essential when comparing a theoretical profile with observations, which cover a finite  radial range  and have finite uncertainties.  It also motivates examining  the full numerical solution rather than relying exclusively on its central expression or asymptotic approximation such as the TF approximation.

To verify the astrophysical viability of the present LHY model, we compare its  predicted circular-velocity profiles with the observed rotation curves of 20 galaxies from the SPARC database \cite{Lelli2016}. 
We look in particular at how the LHY profile reproduces the observed velocity distributions and compare its fitting performance with standard halo prescriptions,  including the NFW \cite{Navarro1996}, the standard  BEC model  \cite{Bohmer2007},
 and the QD2M  findings \cite{Boudjemaa2026}.
The comparison aims  to determine   the  influence of  the full  LHY-EOS  on the resulting rotation-curve fits. The fitting procedure must use consistent  observational data, baryonic contributions, error treatments,  and  parameter-counting conventions 
across the competing models to make the comparison meaningful.
The quality of the statistical fit should be evaluated alongside the physical consistency of the model. Although a lower reduced $\chi$-squared value generally indicates better agreement  with the observational data under a given error model, it does not necessarily imply that a model is statistically preferred when the number of free parameters differs. Information criteria, such as the  Akaike information criterion (AIC) and the Bayesian information criterion (BIC), provide complementary means of comparing competing models by accounting for both goodness of fit and model complexity \cite{Akaike1974,Schwarz1978,Liddle2007}.  As a result, the inferred characteristic halo scales should be viewed as model-dependent estimates, whose interpretation relies on the adopted scaling relations, parameter degeneracies, and assumptions underlying the effective EOS.  

The main purpose of this paper is then to explore whether a self-gravitating  pure-LHY DM  model can provide a useful phenomenological description of galactic DM halos when the full  quantum pressure is retained. 
We derive and analyze the stationary GGPPE, investigate the Jeans-instability criterion \cite{Jean}, limiting properties of the equilibrium solutions, and compare the resulting rotation curves with observations from the SPARC sample.
We  also examine the parameter ranges favored by the adopted fitting procedure and assess the limitations of the model in relation to standard halo descriptions.  
Generally, investigating such an exquisite LHY DM model  is important because it provides an alternative effective description of bosonic DM in which quantum fluctuations, self-gravity and quantum pressure jointly determine halo structure. 
The emerging results may offer a platform for identifying potential observational signatures that distinguish this  scenario from conventional BEC DM and QD2M models, while clarifying the theoretical conditions required for its physical viability. 

The remainder of the paper is organized as follows.  Section ~\ref{GGDM} introduces the effective LHY EOS and the coupled GGPPE. 
In Sec.~\ref{JIns} we present the  linear Jeans instability  analysis. Section~\ref {DimF} is devoted to the dimensionless stationary equations and characteristic halo scales. 
In Sec. ~\ref {Analy} we discuss the analytical limits of the dimensionless GGPPE focusing on the central expansion, asymptotic behavior, and relation to the TF limit, 
We also present its numerical solutions for the full range of the gravitational coupling.
Section~\ref{SPARC} is dedicated to the numerical rotation-curve fits, comparison  with alternative halo profiles,  and to discussion  of  the inferred parameters distributions and statistical model comparison. Section~\ref{concl} summarizes  the main conclusion and outline possible extensions.

\section{generalized Gross-Pitaevskii-Poisson  equations} \label{GGDM}


We consider DM as a self-gravitating Bose fluid in which the dominant local interaction is the repulsive LHY contribution. The essential assumption is that the mean-field interaction is 
entirely removed, and  the dominant self-interaction arises from the beyond mean-field LHY term. The system is then 
described by the following  energy functional:
\begin{equation}\label{Efun}
E [\psi]= \int d^3r \bigg( \frac{\hbar^2}{2m} |\nabla \psi|^2 + \frac{1}{2} m V |\psi|^2 + \frac{2}{5} g_{\text{LHY}} |\psi|^5\bigg),
\end{equation}
where the gravitational potential satisfies the Poisson equation, given by  \cite{Bohmer2007}
$$ \nabla^2 V= 4\pi Gm |\psi|^2.$$
The last term in Eq.~(\ref{Efun}) represents the effective interaction associated with the pure LHY nonlinearity.  This latter originates  from normal and anomalous fluctuations  
which can be  obtained  employing the Hartree-Fock-Bogoliubov theory \cite{Boudjemaa2026,Griffin,Boudj,Boudj1}.
For  $g_{\rm LHY}>0$, this term produces a repulsive nonlinear contribution that  counteracts gravitational attraction and can help establish a finite halo configuration. Unlike the kinetic term (first one), which  accounts for  quantum pressure arising from spatial variation of the wavefunction, the LHY term contributes a local density-dependent interaction energy.

The GGPE can be obtained via $i \hbar\partial \psi/\partial t = \partial E/\partial \psi^*$. This gives 
\begin{equation}\label{GPE}
i\hbar \frac{\partial \psi} {\partial t}= \bigg(-\frac{\hbar^2}{2m} \nabla^2 + m V+  g_{\text{LHY}}  n^{3/2} \bigg) \psi,
\end{equation}
where $n=|\psi|^2$ is the density of the fluid.\\
The pure-LHY chemical potential is 
\begin{equation}\label{EoS}
\mu_{\text{LHY}} = g_{\text{LHY}}  n^{3/2},
\end{equation}
and the corresponding interaction energy reads $E_{\text{LHY}}= \int_0^n \mu_{\text{LHY}} (n') dn'= (2/5) g_{\text{LHY}} n^{5/2}$.
Using the definition $P=n \mu-E$, we obtain for pressure
\begin{equation}\label{EoS1}
P_{\text{LHY}} = \frac{3}{5} g_{\text{LHY}}  n^{5/2}= \frac{3g_{\text{LHY}} }{5 m^{5/2}}  \rho^{5/2},
\end{equation}
where $\rho= m n$ is the LHY-DM density.\\
Thus, the pure-LHY fluid has the polytropic EoS, $P= K \rho^{5/2}$, where $K=3g_{\text{LHY}} / (5 m^{5/2})$, corresponding to $P=K \rho^{1+1/p}$.
This is a polytropic equation with unusual index $p=2/3$ indicating distinctive structural properties of the model.
 
Indeed the use of the above pure-LHY EOS in the present DM model requires some clarification regarding its  microscopic motivation. In a standard single-component dilute Bose gas, the LHY correction originates from quantum fluctuations  beyond the mean-field approximation and scales as $a^{5/2}n^{5/2}$. Therefore, eliminating the mean-field  interaction by taking  $a\rightarrow0$ would simultaneously eliminate the leading  LHY  contribution. A pure LHY-EOS should therefore not be interpreted as the result of simply removing the mean-field term in a single BEC.
As mentioned above, a possible microscopic motivation is provided by a symmetric two-component Bose mixture. For equal masses $m_1=m_2=m$, equal intraspecies interactions $g_{11}=g_{22}=g$, and balanced densities, $n_1=n_2=n$, 
the mean-field contribution can be suppressed  by tuning the  interspecies coupling  close to  $g_{12}=-g$. In this regime, the leading quantum-fluctuation contribution can remain  finite and provide the dominant contribution to the EOS. Such a mechanism  offers a physically motivated route toward an effective LHY-dominated fluid.

The full stationary GGPE can be obtained through $\psi(\mathbf r, t)=\psi(\mathbf r) e^{-i \mu t/\hbar}$, where the ground-state wavefunction can be real. 
For  spherical symmetry,  Eq.~(\ref{GPE}) becomes
\begin{equation}\label{GPE2}
\mu \psi= -\frac{\hbar^2}{2m} \bigg( \frac{d^2\psi}{dr^2} + \frac{2}{r}\frac{d\psi}{dr} \bigg) + m V \psi +  g_{\text{LHY}}  \psi^4,
\end{equation}
Clearly, Eq.~(\ref{GPE2}) contains three competing physical contributions: quantum pressure (first term), gravity (second term), and  LHY repulsion (last term).
The Poisson equation is 
\begin{equation}\label{Pois}
\frac{1}{r^2} \frac{d}{dr}  \left(r^2\frac{dV}{dr} \right) =  4\pi Gm \psi^2.
\end{equation}
Consequently, the full theory is richer than the above TF treatment: the halo can interpolate continuously  between a quantum-supported object and an LHY pressure-supported object.

\section{Jeans instability} \label{JIns}

In this section we investigate the gravitational stability of the pure-LHY Bose fluid by analyzing the evolution of small density perturbations around a homogeneous background.  
The obtained linearized hydrodynamic equations allow us the determine the Jeans instability criterion and examine the competing roles of gravitational attraction, quantum pressure  and LHY-induced repulsion in suppressing the growth of perturbations.

Using Madelung representation: 
\begin{equation} \label {eq9}
\phi( {\mathbf  r},t)=\sqrt{n ( {\mathbf r},t)} \exp [i S (\mathbf r,t)], 
\end{equation}
where $S$ is the phase which is real and related to the superfluid velocity as $v=(\hbar/m) \nabla S$. 
Then, the GGPE (\ref{GPE}) becomes a continuity equation 
\begin{equation}  \label{Hyd1} 
\frac{\partial n} {\partial t} +{\bf \nabla}\cdot (n \mathbf v)=0, 
\end{equation}  
and an Euler-like equation 
\begin{align}  \label{Hyd2} 
m \left[\frac{\partial {\mathbf v}}{\partial t}+(\mathbf v \nabla) \mathbf v\right]  &=-{\bf \nabla} \left (V_Q+ m V  + g_{\text{LHY}} n^{3/2}\right),
\end{align}
where $V_Q=-\left(\hbar^2/2m \right) \nabla^2 \sqrt{n}/ \sqrt{n}$ is the quantum pressure.
This shows that Eq.~(\ref{Hyd2}) strongly depends on the strength of the LHY fluctuations, $g_{\text{LHY}}$. 

We consider a homogeneous background and perturb it according to  $n(\mathbf r,t)= n+ \delta n(\mathbf r,t)$, 
where $\delta n(\mathbf r,t)/n (\mathbf r)\ll1$, and $S=\delta S$.
Linearizing the hydrodynamic equations (\ref{Hyd1})-(\ref{Hyd2}) and retaining only first-order terms, we get
\begin{equation}  \label{hydo3}
\frac{\partial \delta n}{\partial t} + n {\bf \nabla}.\delta \mathbf v=0,
\end{equation}
\begin{equation}  \label{hydo4}
\frac{\partial \delta \mathbf v}{\partial t} = - {\bf \nabla}\delta V- \frac{3 g_{\text{LHY}}}{2m} n^{1/2} {\bf \nabla}\delta n+ \frac{\hbar^2}{4m^2 n} {\bf \nabla} ({\bf \nabla}^2 \delta n).
\end{equation}
and the linearized Poisson equation reads
\begin{equation}  \label{hydo5}
{\bf \nabla}^2\delta V=4\pi G m \delta n.
\end{equation}
Differentiating Eq.~(\ref{hydo3}), taking the divergence of Eq.~(\ref{hydo4}), and using Eq.~(\ref{hydo5}), we obtain
\begin{equation}  \label{hydo6}
 \frac{\partial^2 \delta {n}} {\partial t^2} =4\pi G n m \delta n+ \frac{3 g_{\text{LHY}}}{2m} n^{3/2} {\bf \nabla}^2\delta n- \frac{\hbar^2}{4m^2} {\bf \nabla}^4 \delta n.
\end{equation}
Expanding the solutions of Eq.~(\ref{hydo6}) into  plane waves: $\delta n \propto e^{i ({\mathbf k. \tilde {\mathbf r}}- \omega t)}$, we obtain the pure-LHY Jeans dispersion relation:
\begin{equation}  \label{BogDR}
\omega_k= \sqrt{ \frac {\hbar^2 k^4}{4m^2}+ c_s^2 k^2 - 4\pi mn G}, 
\end{equation}
where 
$$c_s^2= \frac{3g_{\text{LHY}} }{2m} n^{3/2},$$ is the sound velocity.\\ 
Equation (\ref{BogDR}) contains three competing terms, namely: quantum pressure stabilizes the short wavelength, the LHY nonlinearity furnishes an additional stabilizing pressure, 
while gravity destabilizes the long wavelength. The system is therefore unstable whenever $\omega_k^2 <0$.
The Jeans wavenumber satisfies $\omega^2 (J_k)=0$. This gives
\begin{equation}  \label{Jean}
k_J^2 = \frac{2m^2}{\hbar^2}\left(\sqrt{c_s^4+\frac{4\pi G n\hbar^2}{m}}-c_s^2\right).
\end{equation}
The stability criterion is therefore 
$$k<k_J,$$
which means that gravity dominates and the perturbation grows. Once the perturbation becomes nonlinear, $n$ increases and the LHY term becomes increasingly important. 
The system can then settle into the stationary self-gravitating configuration governed by the GGPPE.

Assuming that the quantum term is negligible at the Jeans scale. Then Eq.~(\ref{Jean}) reduces
\begin{equation}  \label{Jean1}
k_{J,\rm LHY}^2 = \frac{8\pi G m^2}{3 g_{LHY}\sqrt n},
\end{equation}
which is the Jeans scale of the pure-LHY dominated regime.\\
In the quantum regime, where the LHY contribution is  negligible, one has
\begin{equation}  \label{Jean2}
k_{J,Q}^2 = \left(\frac{16\pi G m^3 n}{\hbar^2}\right)^{1/2}.
\end{equation}
The quantum-LHY crossover occurs at
\[
\frac{\hbar^2 k^4}{4m^2} \sim  \frac{3g_{\text{LHY}} }{2m} n^{3/2} k^2,
\]
its corresponding wavenumber reads 
\begin{equation}  \label{Jean3}
k_{J,Q\text{-}\rm LHY}^2 = \frac{6 m g_{LHY} n^{3/2}}{\hbar^2}.
\end{equation}
This is indeed important since it lets us to classify the instability.
For $k_J \gg k_{J,Q-{\rm LHY}}$, quantum pressure dominates over the LHY term in establishing stability, while for $k_J \ll k_{J,Q-{\rm LHY}}$,  the opposite situation holds.

For unstable modes, $\omega_k^2<0$, we can write $\omega=i\Gamma$, which gives for the Jeans growth rate:
\begin{equation}  \label{Jean4}
\Gamma_k^2 = 4\pi G m n - c_s^2k^2 - \frac{\hbar^2k^4}{4m^2}.
\end{equation}
Hence $\delta n \propto e^{\Gamma t} $.
At  $k=0$, $\Gamma_0=\sqrt{4\pi Gnm}$, which is nothing else than the usual gravitational growth rate. 
As $k$ increases, the LHY and quantum terms suppress the growth until $\Gamma^2_{k_J}=0$.

We also define the Jeans mass as $M_J = 4 \pi m n R_J^3/3$, where $R_J =\pi/k_J$. Then substituting Eq.~(\ref{Jean}), one has 
\begin{align}  \label{Jean5}
M_J&= \frac{4 \pi^4 m n}{3}  \bigg\{ \frac{2m^2}{\hbar^2} \bigg[ \sqrt{  \left(\frac{3g_{\text{LHY}} }{2m} n^{3/2} \right)^2 + \frac{4\pi Gn \hbar^2}{m} } \nonumber \\
&-\frac{3g_{\text{LHY}} }{2m} n^{3/2} \bigg] \bigg\}^{-3/2}.  
\end{align}
In the quantum regime, $g_{\rm{LHY}} \rightarrow 0$, we obtain for Jeans mass upon inserting (\ref{Jean2}):
\begin{align}  \label{Jean6}
M_{J,Q}= \frac{\pi^{13/4} \hbar^{3/2} n^{1/4}}{ 6 G^{3/4} m^{5/4}}.  
\end{align}
In the LHY-dominated regime $V_Q \rightarrow 0$, the use of Eq.~(\ref{Jean1}) gives:
\begin{align}  \label{Jean6}
M_{J,\rm LHY} = \frac{\sqrt6\,\pi^{5/2}}{8}\,\frac{g_{LHY}^{3/2}\,n^{7/4}}{G^{3/2}\,m^{2}}.
\end{align}
Thus, increasing the LHY coupling strongly increases the characteristic mass scale below which gravitational collapse is suppressed.
An important remark is that $M_J \propto g_{\rm {LHY}}^{3/2}$ since $R_J \propto g_{\rm {LHY}}^{1/2}$, which provides a direct connection between the linear Jeans instability 
and the nonlinear halo scale.

\begin{figure}
	\includegraphics[scale=0.75]{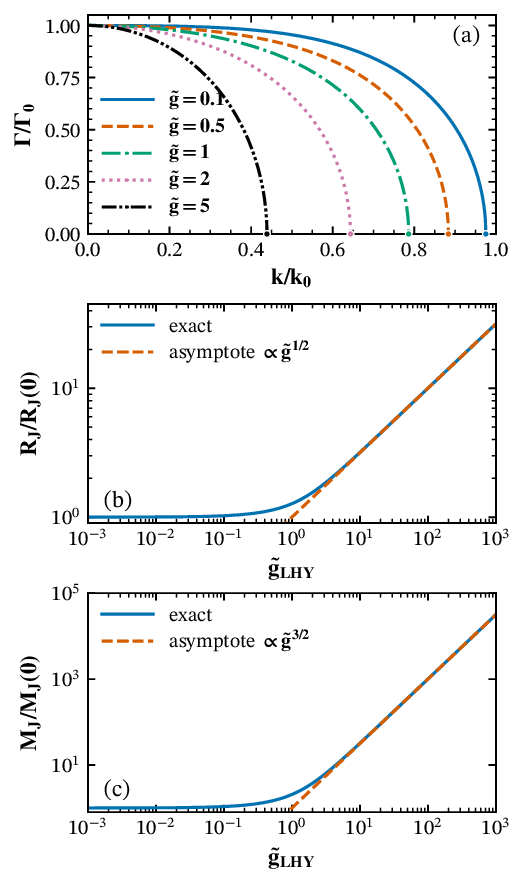}
		\caption{ (a) Dimensionless Jeans growth, $\Gamma/\Gamma_0$, as a function of $k/k_0$ for  several values of  the dimensionless LHY coupling $ \tilde g_{\rm LHY}$; dots mark the Jeans cutoff  $k_J/k_0$ beyond which perturbations are unstable. 
(b) Jeans radius, $R_J/R_J(0)$ and  (c)  Jeans mass,  $M_J/M_J(0)$, as a function of  $\tilde g_{\rm LHY}$ at fixed density. Dashed lines show the large-coupling asymptotes  $\tilde g_{\rm LHY}^{1/2}$ and $\tilde g_{\rm LHY}^{3/2}$.
Here $k_0=(16\pi G m^{3}n/\hbar^{2})^{1/4}$, $\Gamma_0=\sqrt{4\pi G m n}$, $R_J(0)=\pi/k_0$ and $M_J(0)=4\pi m n R_J^{3}(0)/3$.}
\label{fig1} 
\end{figure}

To quantify the role of the LHY term in the linear regime, we introduce the dimensionless coupling
\begin{equation}
\tilde g_{\rm LHY}\equiv\frac{6\,m\,g_{\rm LHY}\,n^{3/2}}{\hbar^{2}k_0^{2}}
=\frac{k_{J,Q\text{-}{\rm LHY}}^{2}}{k_0^{2}}
=\frac{3\,g_{\rm LHY}\,n^{3/2}}{\hbar \Gamma_0},
\end{equation}
where $k_0=k_{J,Q}$. In terms of  $x=k/k_0$, the growth rate of Eq.~(\ref {Jean4}) turns out to be given as:
$\Gamma^{2}/\Gamma_0^{2}=1-\tilde g_{\rm LHY}x^{2}-x^{4}$, and the Jeans  cutoff reads $k_J/k_0=\left[(\sqrt{\tilde g_{\rm LHY}^{2}+4}-\tilde g_{\rm LHY})/2\right]^{1/2}$.
The Jeans mass radius and mass then follow as: 
$R_J/R_J(0)=\left[(\tilde g_{\rm LHY}+\sqrt{\tilde g_{\rm LHY}^{2}+4})/2\right]^{1/2}$, and  $M_J/M_J(0)=\left[R_J/R_J(0)\right]^{3}$ at fixed density.

Figure~\ref{fig1} (a) depicts the  growth rate as a function of  $k/k_0$. 
For every value of $\tilde g_{\rm LHY}$ the growth rate  equals the free-fall value $\Gamma_0$ at $k=0$, so  the LHY repulsion does not slow  the long-wavelength modes.
It acts only on shorter wavelengths: for small $k$,  $\Gamma\simeq\Gamma_0\left(1-\tilde g_{\rm LHY}k^{2}/2k_0^{2}\right)$, and increasing  $\tilde g_{\rm LHY}$  narrows  the  unstable band.  
The cutoff wavenumber  decreases from $k_J/k_0\simeq 0.975$ for $\tilde g_{\rm LHY}=0.1$  to $0.884$,  $0.786$, $0.644$ and $0.439$ for  $\tilde g_{\rm LHY}=0.5$, $1$, $2$ and $5$, respectively. 
Since both pressure terms are positive, there is no fastest growing mode at finite  $k$: the  growth rate decreases monotonically, and all modes with  $k>k_J$ are stable.

Figure~\ref{fig1} (b)  and (c) show the Jeans radius and mass relative to their quantum-pressure values.  
For   $\tilde g_{\rm LHY}\lesssim0.1$ both are essentially  unchanged ($R_J/R_J(0)\simeq1.03$ at $\tilde g_{\rm LHY}=0.1$),  indicating that quantum pressure sets the Jeans scale. 
The crossover to LHY support occurs at $\tilde g_{\rm LHY}\sim1$, where $k_{J,Q\text{-}{\rm LHY}}\sim k_0$, corresponding to  $R_J/R_J(0)\simeq1.27$ and $M_J/M_J(0)\simeq2.06$.
For  $\tilde g_{\rm LHY}\gg1$, the curves approaches the LHY-dominated power laws  $R_J\propto\tilde g_{\rm LHY}^{1/2}$  and  $M_J\propto\tilde g_{\rm LHY}^{3/2}$ of Eqs.~(\ref{Jean1}) and (\ref{Jean6}) reaching $R_J/R_J(0)\simeq10$ and 
$M_J/M_J(0)\simeq10^{3}$ at $\tilde g_{\rm LHY}=100$. 
The LHY repulsion thus raises the minimum mass that can collapse  gravitationally and the effect grows steeply  with the coupling. Moreover,  in the LHY-dominant regime  $R_J\propto n^{1/4}$, a  much weaker density  dependences than the 
$R_J\propto n^{-1/4}$  of the purely quantum regime (due to  $k_{J,Q}^{2}\propto n^{1/2}$ ). For a standard  contact-interaction BEC,   $R_J$  is density independent. 

Since $\tilde g_{\rm LHY}\propto g_{\rm LHY}\,n$ at fixed particle parameters, denser regions such as cores are driven toward the LHY-dominated regime ($\tilde g_{\rm LHY}\gg1$), whereas diffuse regions remain quantum-pressure dominated. 
This is consistent with the picture of the nonlinear halo, in which  the LHY term supports  the core and quantum pressure  shapes the outer tail. 
We emphasize that the analysis is linear and static, it neglects the cosmological  expansion  and baryons,  and the numerical values depend on the boson mass and the density through  
$\tilde g_{\rm LHY}$.
We can infer that the LHY interaction plays a stabilizing role in a self-gravitating quantum fluid. 
These results provide a connection between the linear stability analysis and the subsequent study on nonlinear LHY-supported DM halos.

\section{Dimensionless formulation} \label{DimF}

Here, we  introduce the dimensionless formulation of the coupled GGPPE to facilitate the numerical analysis of the halo profiles and to identify the key parameters controlling the interplay of quantum pressure, LHY and self-gravity.

It is convenient to define the wavefunction as
\begin{equation}\label{WF}
\psi(r)= \sqrt{n_0} f(\tilde r),  
\end{equation}
where $n_0= |\psi(0)|^2$ is the central density.\\
We define the corresponding quantum-LHY length as:
\begin{equation}\label{leg}
\ell_{\text{LHY}}= \frac{\hbar} {\sqrt{2m g_{\text{LHY}}  n_0^{3/2} }}.
\end{equation}
Then, 
$\tilde r= r/\ell_{\text{LHY}}$, $V( r)= \left(g_{\text{LHY}} n_0^{3/2}/m \right)  \tilde V(r)$, and $\tilde \mu= \mu/ \left(g_{\text{LHY}} n_0^{3/2}\right)$.
Therefore, the stationary equations reduce to: 
\begin{equation}\label{SGPE2}
f^{''}+\frac{2}{\tilde r} f^{'}= (V+f^3- \tilde \mu) f,
\end{equation}
\begin{equation}\label{Pois1} 
V^{''}+ \frac{2}{\tilde r} V^{'}= \lambda f^2,
\end{equation}
where 
\begin{equation}\label{lamd}
\lambda= \frac{2\pi G m \hbar^2}{g_{\text{LHY}}^2 n_0^2},
\end{equation}
which cannot be viewed only as simply the quantum-to-LHY ratio, but it measures gravitational coupling in the particular central-density scaling chosen above.\\
The boundary conditions are: 
$$ f(0)=1, \;\;\;\;  f^{'}(0)=0,\;\;\;\; V^{'} (0)=0,$$
together with localization 
$$ f(\tilde r= \infty) \rightarrow 0.$$
One may choose the gauge $ V (0)=0$.
The chemical potential $ \tilde \mu$ is then an eigenvalue fixed by the localization condition.

Importantly, this model predicts a family of halo profiles.\\
The normalized density is 
\begin{equation}\label{Ndens}
\rho= \rho_0 f_{\lambda}^2 (\tilde r).
\end{equation}
The dimensionless enclosed mass is 
\begin{equation}\label{Nmas}
\mathcal M_{\lambda} (\tilde r) = \int_0^{\tilde r}  \tilde r'^{2} f_{\lambda}^2 (\tilde r') d\tilde r' .
\end{equation}
Thus, 
\begin{equation}\label{Nmas1}
M= 4 \pi m n_0 \ell_{\text{LHY}}^3 \mathcal M_{\lambda} (\tilde r).
\end{equation} 
The circular velocity follows from  $v_{\text{tg}}^2 (\tilde r)= G M(\tilde r)/\tilde r$, yielding 
\begin{equation}\label{NVel1}
v_{\text{tg}}^2 (\tilde r)= 4 \pi G m n_0 \ell_{\text{LHY}}^2 \frac{\mathcal M_{\lambda} (\tilde r)}{\tilde r}.
\end{equation}
Therefore, every solution provides a complete set of universal dimensionless functions  namely, density profile $f_{\lambda}^2 (\tilde r)$, mass profile $\mathcal M_{\lambda} (\tilde r)$,
and rotation curves  $\mathcal M_{\lambda} (\tilde r)/\tilde r$.

\section{Analytical and numerical solutions} \label{Analy}

In this section, we study analytical limits of the dimensionless GGPPE describing  pure-LHY DM halos. 
Although a general closed-form solution is unavailable, the central ($\tilde r \rightarrow 0$) and  asymptotic  ($\tilde r \rightarrow \infty$) behaviors can be derived  explicitly. 
The central expansion characterizes the inner density profile, while the large distance radius limit reveals how quantum pressure controls the halo density decay. 
This may provide physical insight into the halo structure and useful consistency checks for the numerical solutions. 

\subsection{Near the halo center} 

Based on the above condition, we can write 
\begin{equation}\label{SGPE3}
f (\tilde r)= 1+ a  \tilde r^2+ b \tilde r^4+ \cdots,
\end{equation}
and 
\begin{equation}\label{Pois2}
V(\tilde r)= c \tilde r^2+ d \tilde r^4+ \cdots.
\end{equation}
Substitution into Eqs.~(\ref{SGPE2}) and (\ref{Pois1}) gives: $a= (1-\tilde \mu)/6$, and $c=\lambda/6$. Thus,   
\begin{equation}\label{SGPE4}
f (\tilde r)= 1+ \frac{1-\tilde \mu }{6} \tilde r^2+ \cdots
\end{equation}
For a monotonically decreasing ground-state, we should have $\tilde \mu >1$.\\
Likewise, the gravitational potential takes the form: 
\begin{equation}\label{Pois3}
V(\tilde r)= \frac{\lambda}{6} \tilde r^2+ \cdots
\end{equation}
Hence for nodeless ground-state, the density reads
\begin{equation}\label{Ndens1}
\rho \simeq \rho_0\left[1+\frac{1-\tilde\mu}{3}\,\tilde r^{2}\right],
\end{equation}
Interestingly,  Eq.~(\ref{Ndens1}) shows that the center is cored rather than cuspy.\\
The enclosed mass can be immediately obtained from Eq.~(\ref{Nmas1}) as:
\begin{equation}\label{Nmas2}
M(\tilde r)=M_0(\tilde r)+\frac{4\pi}{15}\,\rho_0\,\ell_{\rm LHY}^{3}\,(1-\tilde\mu)\,\tilde r^{5},
\end{equation} 
where $ M_0(\tilde r)=4\pi\,\rho_0\,\ell_{\rm LHY}^{3}\,\tilde r^{3}/3$  is the mass of a constant-density sphere.\\
As a result, the rotation curve (\ref{NVel1}) becomes
\begin{equation}\label{NVel1}
v_{\rm tg}^{2}(\tilde r)=v_{\rm tg0}^{2}(\tilde r)
+\frac{4\pi G\rho_0\,\ell_{\rm LHY}^{2}}{15}\,(1-\tilde\mu)\,\tilde r^{4}.
\end{equation}
Since $\tilde\mu>1$, the quartic correction in Eq.~(\ref{NVel1}) is negative, then  $v_{\rm tg}(\tilde r)<v_{\rm tg0}(\tilde r)$; the rotation curve rises more slowly than the linear, constant-density result, with the deviation growing as  $\tilde r^{2}$ (valid for  $\tilde r\ll1$).
This  leads to an important observational prediction, $v_{\text{tg}0} \propto r$ for $r \rightarrow 0$:
\[
v_{\rm tg0}^{2}(\tilde r)=\frac{4\pi G\rho_0\,\ell_{\rm LHY}^{2}}{3}\,\tilde r^{2}.
\]
The rotation curve increases linearly in the central region because the LHY halo possesses a finite central density.
Thus, the pure-LHY halo inherently produces a cored rotation curve, rather than the divergent central  behavior associated with cuspy density profile.

\subsection{Asymptotic quantum tail} 
At large distances, $f\ll1$, and the LHY nonlinearity becomes negligible. The equation asymptotically reduces to 
\[
f^{''}+\frac{2}{\tilde r} f^{'} \simeq   \kappa^2 f,
\]
where $\kappa^2= V (\infty)- \tilde \mu >0$.\\
The decaying solution behaves as:
\begin{equation}\label{SGPE5}
f (\tilde r) \sim \frac{A}{\tilde r} e^{-\kappa \tilde r }.
\end{equation}
Therefore, the LHY-DM density reads 
\begin{equation}\label{Ndens2}
\rho \sim \frac{A^2}{\tilde r^2} e^{-2\kappa \tilde r },
\end{equation}
pointing out that quantum LHY halo does not have  an abrupt finite boundary of the TF solution.
Instead, quantum pressure smooths the LHY edge into an exponentially decaying outer atmosphere. \\
The mass outside radius $r$ is 
\begin{equation}\label{Nmas3}
M_{\infty}- M(r)=  4\pi \rho_0 \ell_{\text{LHY}}^3 \int_{\tilde r}^{\infty} \tilde r^2 f^2(\tilde r) d\tilde r.
\end{equation} 
Using the asymptotic form: $\tilde r^2 f^2( \tilde r) \simeq A^2 e^{-2\kappa \tilde r}$, we get
\begin{equation}\label{Nmas4}
 M(r) \simeq  M_{\infty}-  \frac{2\pi \rho_0 \ell_{\text{LHY}}^3 A^2}{\kappa} e^{-2\kappa  r/\ell_{\text{LHY}}}.
\end{equation} 
We see that the total mass converges exponentially. At sufficiently large radii, $ M(r) \rightarrow  M_{\infty}$.\\
The rotation velocity is then 
\begin{equation}\label{NVel2}
v_{\text{tg}} (\tilde r) \simeq \sqrt { \frac{ G  M_{\infty} } {r} },  \:\:\;\;\; r \rightarrow \infty,
\end{equation}
which is the standard Keplerian falloff. So, $v_{\text{tg}} \propto r^{-1/2}$ which means that the pure LHY halo does not induce an asymptotically flat rotation curve.
Such a flat rotation curve can nevertheless occur over an intermediate radial interval.

\subsection{Thomas-Fermi limit}  \label{ASTF}
\begin{figure}
	\includegraphics[scale=0.75]{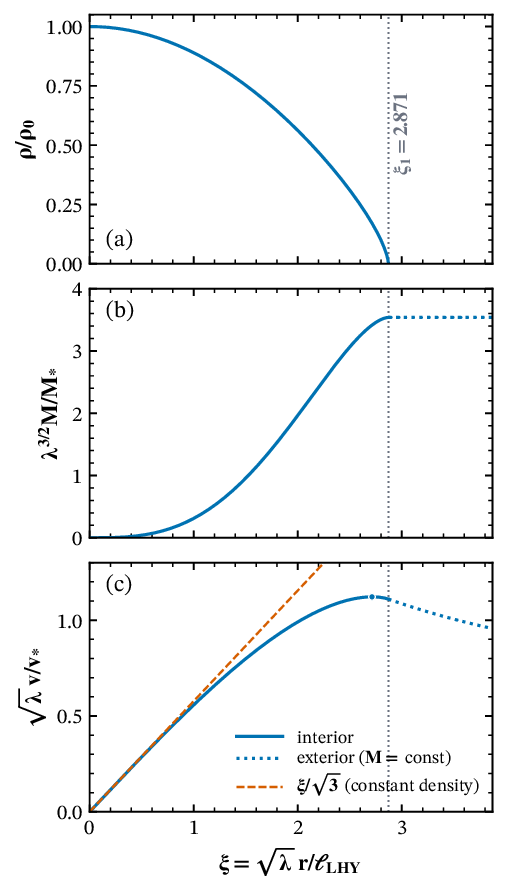}
		\caption{ TF-limit of the LHY halo (Lane-Emden solution with index p=2/3, Eq.~(\ref{LEE})) as a function of $\xi$. (a) Density profile, $\rho (\xi)/ \rho_0$, which vanishes at the halo edge $\xi_1=2.871$ (vertical dotted line).
		(b) Enclosed mass profile, $\lambda^{3/2} M(\xi)/ (4\pi \rho_0 \ell_{\rm LHY}^3$, it saturates at 3.54 at $\xi_1$. (c) Tangential velocity, $\sqrt{\lambda} v_{\rm tg} /v_*$, with $v_{*}=\sqrt{4\pi G\rho_0}\,\ell_{\rm LHY}$. 
		The solid line is the interior solution and the dotted line the exterior Keplerian fall-off $\xi^{-1/2}$; the dashed line is the constant-density core $\xi/\sqrt{3}$. The velocity peaks at 1.12 at $\xi=2.71$.} 
\label{fig2} 
\end{figure}

When quantum pressure is negligible $\nabla^2 f \ll f^4$, we have $ V+f^3= \tilde \mu$. Defining $\theta=f^3$, and $\xi= \sqrt{\lambda} \tilde r$, yields
\begin{equation}\label{LEE}
\frac{1}{\xi^2} \frac{d}{d\xi} \left(\xi^2 \frac{d \theta}{d\xi} \right) + \theta^{2/3}=0,  
\end{equation}
with $\theta(0)=1$, and $\theta'(0)=0$.
This is the Lane-Emden equation with index $p=2/3$, which has no known elementary analytical solution. Thus, Eq.~(\ref{LEE}) should be solved numerically.
The results for density profile, enclosed mass profile, and tangential velocity are shown in Fig.~\ref{fig2}.

Figure \ref{fig2} (a) demonstrates that the density $\rho (\xi)/ \rho_0$ decreases smoothly and monotonically with $\xi$. 
The decrease is relatively slow near the center and becomes progressively steeper toward the outer region. Importantly, the density reaches $\rho (\xi)=0$  at a finite radius. 
As a result, our TF-LHY solution describes a finite-size halo, $\xi  \simeq 2. 871$, with a dense central region. 
The absence of a central cusp is also visible; the profile is smooth  at $\xi=0$, with its maximum at the center.
We see from Fig. \ref{fig2} (b) that the mass grows slowly in the central region, and then much more rapidly around $\xi =1.5-2.3$, before progressively flattening as $\xi$ approaches the halo boundary.
At the edge, the dimensionless enclosed mass is $\lambda^{3/2}M_h\simeq 3.5 (4\pi \rho_0 \ell_{\rm LHY}^3)$.

The flattening near the edge means that when the density approaches zero, there is essentially no additional mass to accumulate outside the halo.
The  tangential velocity is plotted in Fig. \ref{fig2} (c).  
It is apparent that it  initially rises almost linearly with radius, $\sqrt{\lambda}\,v_{\rm tg}/v_*\simeq\xi/\sqrt3$,  as expected for a constant-density core, and the slope then progressively decreases.
Near the outer edge, $\xi \simeq 2-2.5$,  the velocity  flattens, and it  reaches a maximum $\sqrt{\lambda}\,v_{\rm tg}/v_*\simeq1.12$ at $\xi\simeq2.71$, slightly inside the edge.
Since the TF halo has a sharp boundary  and  a finite mass, the velocity at the edge is  $\simeq1.11$ and beyond $\xi_1$ the exterior curve follows the Keplerian fall-off $v_{\rm tg}\propto\xi^{-1/2}$ [see dotted line in Fig.~\ref{fig2}(c)]. 
The rotation curve therefore has a broad maximum of nearly constant velocity around $\xi\simeq2-3$, followed by a Keplerian decline, rather than an asymptotically flat plateau.
Compared with standard BEC DM \cite{Bohmer2007,Chavanis2011}, and with our QD2M model  \cite{Boudjemaa2026},  the present pure-LHY self-gravitating fluid, in which LHY repulsion alone counterbalances gravity, leads to finite, cored halo with a characteristic mass and rotation-curve profile.

\subsection{Halo profiles at finite $\lambda$} \label{lambda}

The above analytic expansions describe the halo only near the center, in the tail, and in the TF limit. 
To cover the full range of the gravitational coupling $\lambda=8\pi G m^{2}\rho_0\ell_{\rm LHY}^{4}/\hbar^{2}$, we solve the full (non-TF) Eqs.~(\ref{SGPE2}) and (\ref{Pois1})  numerically.
These equations  are integrated on  $\tilde r\in[R_0,R_{\max}]$ with $R_0=10^{-4}$, starting from the series $f=1+(1-\tilde\mu)\tilde r^2/6$, $V=\lambda\tilde r^2/6$ and their derivatives,  which removes the $2/\tilde r$  singularity and imposes $f(0)=1$, $f'(0)=V(0)=V'(0)=0$.
The eigenvalue $\tilde\mu>1$ is fixed by requiring the nodeless solution to decay,  $f(R_{\max})=0$; $R_{\max}$  is increased until $\tilde\mu$ and $M_\lambda$ change by less than $10^{-6}$.
The problem is solved as a boundary-value problem (collocation, tolerance $10^{-8}$) with $\tilde\mu$ as  an unknown parameter and continuation in $\lambda$, seeded at $\lambda=1$ 
by shooting solution (\textsc{DOP853}, bisection on  $\tilde\mu$) \cite{Hairer1996,Ascher1995}.
Shooting alone is ill-conditioned for  $\lambda\lesssim0.1$  because the solution is exponentially sensitive to $\tilde\mu$ at large $\tilde r$. 
The enclosed mass obtained by quadrature agrees with $\tilde r^{2}V'/\lambda$ to $10^{-6}$ and for $\lambda\to0$  the solutions reproduce the Lane-Emden values $\xi_1=2.8713$,  $m(\xi_1)=3.5387$.

\begin{figure}
	\includegraphics[scale=0.45]{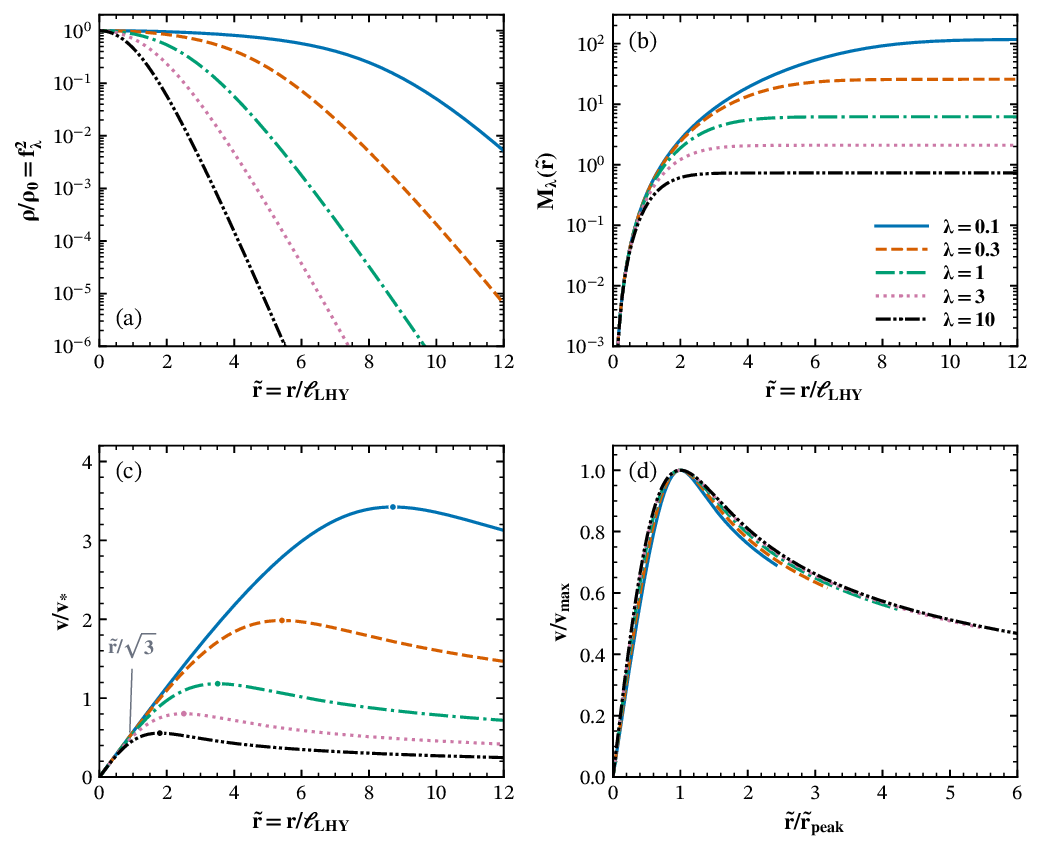}
		\caption{ Dependence of the pure-LHY halo structure on the gravitational coupling $\lambda=8\pi G m^{2}\rho_0\ell_{\rm LHY}^{4}/\hbar^{2}$, obtained by solving the full dimensionless GGPPE (\ref{SGPE2}) and (\ref{Pois1}) for $\lambda=0.1$ (solid), $0.3$ (dashed), $1$ (dash-dotted), $3$ (dotted) and  $10$ (dash-dot-dotted). 
		(a)~ Density profile $\rho/\rho_0=f_\lambda^{2}$ in logarithmic scale.   
		(b)~ Enclosed mass $\mathcal{M}_\lambda(\tilde r)$ in logarithmic scale, which saturates at $\mathcal{M}_\infty(\lambda)$
		(c)~Tangential velocity $v/v_{*}$; dots mark the maximum and thin dotted universal inner behavior $v/v_{*}=\tilde r/\sqrt{3}$.
		(d)~ The same curves normalized to their maximum $v/v_{\rm max}$ versus  $\tilde r/\tilde r_{\rm peak}$. Here $\tilde r=r/\ell_{\rm LHY}$. }
\label{fig22} 
\end{figure}

Figure~\ref{fig22} shows how halo structure depends on the dimensionless coupling $\lambda$, which controls the strength of self-gravity relative to the LHY pressure.
The profiles are obtained from the full eigenvalue problem, including the quantum pressure term, and the TF limit serves only as a check.
For $\lambda=0.1$ the total mass, $\mathcal{M}_\infty\lambda^{3/2}\simeq3.73$, lies within about 5.4\% of the Lane-Emden $3.54$ obtained for $p=2/3$.
In all cases the center is cored, $\rho\simeq\rho_0\bigl[1+(1-\tilde\mu)\,\tilde r^{2}/3\bigr]$, and rotation curve rises as $v=v_{*}\tilde r/\sqrt3$ independently of $\lambda$ 
[see Fig.~\ref{fig22} (c)]  ($\lambda$  only determines the radius at which the curves depart from this universal inner behavior).
Increasing $\lambda$ makes the halo more compact; the half-mass radius decreases from $\tilde r_{1/2}\simeq6.3$  at $\lambda=0.1$ to $1.3$ at $\lambda=10$, 
the total mass $\mathcal{M}_\infty$  which is finite because  $\mathcal{M}_\lambda(\tilde r)$ saturates  [see Fig.~\ref{fig22}(b)], 
decreases by more than two orders of magnitude ( from 118 to 0.73), and the chemical potential  grows from $\tilde\mu\simeq 1.03$ to  $3.30$, consistent with $\tilde\mu-1\simeq0.34\lambda$  
at small $\lambda$. In such a case the density falls off nearly like a TF halo, whereas for large $\lambda$, the kinetic term smooths the edge into an exponentially decaying tail $f\propto e^{-\kappa\tilde r}/\tilde r$ which steepens with $\lambda$ [see Fig.~\ref{fig22}(a)].
Because the mass is finite, the rotation curve reaches a maximum,  for instance  $v_{\max}/v_*\simeq 3.4$ at $\tilde r\simeq 8.7$  for $\lambda=0.1$, and 
$v_{\rm max}/v_{*}\simeq0.56$ at $\tilde r\simeq1.8$ for $\lambda=10$, and then declines toward the Keplerian behavior $v\propto\tilde r^{-1/2}$ [see Fig.~\ref{fig22} (c)]. Once rescaled by the position  and height of maximum, the five curves collapse onto each other to within a few percent 
with larger $\lambda$ giving slightly slower decline  beyond the peak [see Fig.~\ref{fig22} (d)].  
The observable shape of a rotation curve is therefore fixed mainly by the pair  $(\tilde r_{\rm peak},v_{\rm max})$, that is, by  $\rho_0$ and $\ell_{\rm LHY}$, and 
depends only weakly on $\lambda$.  As a consequence, since $\lambda\propto m^{2}\rho_0\ell_{\rm LHY}^{4}$, rotation-curve fits alone do not constrain the boson mass; $\rho_0$ and 
$\ell_{\rm LHY}$ are the physically meaningful fit parameters (see below).

\section{Application of LHY-DM halos to SPARC galactic rotation curves}  \label{SPARC} 

To model galactic rotation curves, we  use the solution of Sec.~\ref{lambda},  (same solver and eigenvalue condition). 
For each $(\rho_0,\ell_{\rm LHY})$ the dimensionless profile is taken from a precomputed table of $\ln(M_\lambda/\tilde r^{3})$
on a $56\times400$ grid in $(\ln\lambda,\ln\tilde r)$ covering $0.05\le\lambda\le4000$ (bicubic interpolation; the maximum relative error
in $v$ is $1.5\times10^{-4}$ against direct solutions at ten off-grid values of $\lambda$), and $v_{\rm DM}^2=4\pi G\rho_0\ell_{\rm LHY}^2M_\lambda(\tilde r)/\tilde r$. For $\lambda\gtrsim5$ the profile changes only by a rescaling of
$(\tilde r_{\rm peak},v_{\max})$ (shape differences $\lesssim2\%$, Sec.~\ref{lambda}), so $\rho_0$ and $\ell_{\rm LHY}$ act as two effective scale parameters.
In the fits $\lambda$ is set to the table value $\lambda=4000$ for 18 galaxies and at $\lambda=0.05$ for DDO~168 and UGCA~442;
the boson mass is \emph{not} fixed. The implied masses, $m_{\rm eff}=m\sqrt{\lambda_{\rm used}/\lambda(\rho_0,\ell_{\rm LHY};m)}$, span
$8\times10^{-26}$--$5\times10^{-23}$~eV (median $5\times10^{-24}$~eV) for the 18 large-$\lambda$ galaxies.

The LHY halo interpolates between known limits. For small $\lambda$, it approaches the TF configuration of Sec.~\ref{ASTF}, and for large $\lambda$, it approaches the quantum pressure-supported soliton of the Schr\"odinger-Poisson system.
Most of the fitted galaxies lie in the latter regime, where the rotation curves constrain the two scale parameters  ($\rho_0, \ell_{\rm LHY}$) but not the interaction strength separately. 
DDO 168 and UGCA 442 lie in the TF regime, where the LHY nonlinearity directly shapes the halo.

The physical halo parameters including central density, $\rho_0$, and quantum scale radius $\ell_{\rm LHY}$, are optimized against observational SPARC galaxy kinematics 
using a Levenberg-Marquardt nonlinear least-squares algorithm to minimize the reduced $\chi_{\nu}^2$ statistic across baryonic and DM potential components \cite{Press2007, Liddle2007}:
$$
\chi_{\nu}^2= \frac{1}{N-K} \sum_i^N \bigg[ \frac{v_{\rm obs}  (r_i) -v_{\rm total} ( r_i; \rho_0, \ell_{\rm LHY})}{\sigma_i}\bigg]^2,
$$
where $\sigma_i$ is the observational uncertainty  associated with the observed rotation velocity, $v_{\rm obs}$, at the radial point $r_i$, and $K$ is the number of fitted parameters. 
The total velocity can be written as:
$$
v_{\rm total} ( r)= \sqrt{v_{\rm gas}^2(r)+ \Upsilon_{\rm disk} v_{\rm disk}^2+  \Upsilon_{\rm bulge} v_{\rm bulge}^2+ v_{\rm DM}^2(r)},
$$
where $v_{\rm gas} (r)$ is the gas component of the velocity, $v_{\rm disk}$ is the stellar disk component velocity,  $ v_{\rm bulge}$ is the bulge contribution, and $v_{\rm DM}$ is the DM part velocity, with $\Upsilon_{\rm disk}$ and $\Upsilon_{\rm bulge}$ being the stellar mass-to-light ratios for the disk and the bulge, respectively. 
The gas and stellar terms enter as signed quantities, $v_i^2\to v_i|v_i|$, as in SPARC. The  stellar mass-to-light ratios are fixed as the standard 3.6\,$\mu$m values 
 $\Upsilon_{\rm disk}=0.5$ and $\Upsilon_{\rm bulge}=0.7\,M_\odot/L_\odot$ and  are not fitted, so  the only free parameters are $(\rho_0,\ell_{\rm LHY})$, i.e. \ $K=2$.
The bulge term is set zero only for galaxies without a bulge.


\subsection{Observational fits and comparative analysis of DM Halo models}
\label{subsec:comparative_analysis}

To check  the physical viability of our LHY DM halo, we benchmark its performance against three prominent theoretical halo profiles across our 20-galaxy SPARC sample, including 
the cuspy cold DM (NFW) profile  \cite{Navarro1996,Press2007}, the standard BEC DM  governed by only the two-body contact interactions  \cite{Bohmer2007,Chavanis2011}, 
and  our recently proposed QD2M model formed by ultradilute binary BEC mixtures   \cite{Boudjemaa2026}.

\begin{figure*}
	\includegraphics[scale=0.55]{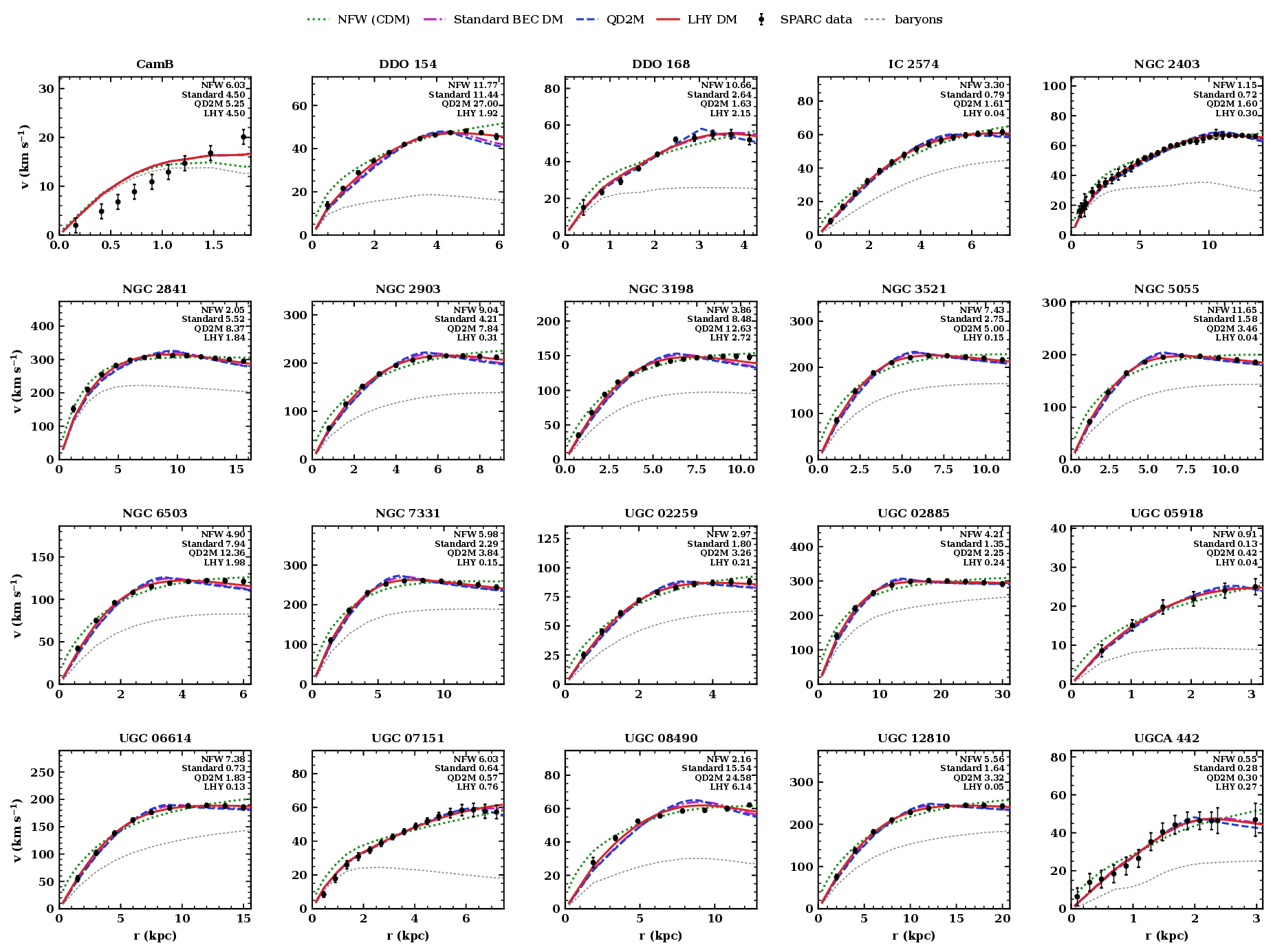}
		\caption{ Rotation curves for 20 SPARC galaxies. The observed data points are shown with their respective uncertainties (error bars).
		Curves show the total velocity (baryon plus DM halo) for the pure-LHY  (red solid), QD2M (blue dashed), standard BEC DM (magenta dash-dotted), and NFW ( green dotted); 
		the grey dashed curve represents the baryonic contribution, with $\Upsilon_{\rm disk}=0.5$, and $\Upsilon_{\rm bulge}=0.7$ fixed.  
		Panel labels list the reduced $\chi^2_\nu$  of each  model, as in Table~\ref {tab:chi2_comparison}. }
\label{fig3} 
\end{figure*}

Figure~\ref{fig3}  illustrates the total velocity fits for all four models alongside the observed SPARC kinematics. The models exhibit fundamental structural and kinematic distinctions across different radial regimes.
First, the NFW profile exhibits a characteristic $1/r$  central density divergence (cusp), which yields a steeper inner velocity rise  ($v \propto r^{1/2}$) \cite{Press2007}. 
As a result, NFW systematically overpredicts central rotation speeds in DM dominated dwarf and low-surface-brightness (LSB) galaxies  (e.g., CamB, DDO 154, IC 2574),
leading to $\chi^2_\nu$ up to 11.8 (DDO~154), with a sample mean $\langle \chi^2_{\nu} \rangle=5.38$ (see Table~\ref{tab:chi2_comparison}).
In contrast,  the quantum fluid models (BEC DM, QD2M, and LHY DM) establish a smooth, consistent-density central core $\rho_0$ stabilized by quantum pressure and/or repulsive interaction. 
The cored geometry produces a strictly linear velocity rise  ($v \propto r$), clearly resolving the observed cusp-core problem without requiring a fine-tuned baryonic feedback.
Second, standard BEC halos in TF regime terminate sharply at a finite radius $R_h = \pi/k$, producing a pronounced kink in  $v_{\text{DM}}(r)$ near the halo boundary. 
The QD2M droplet model forms a self-bound  configuration with a flat-top central profile and a relatively sharp surface drop-off  \cite{Boudjemaa2026}  causing the rotation curve to transition abruptly toward the outer plateau.
In contrast, the LHY DM  model incorporates full quantum wavefunction gradient term ($\nabla^2 f$), which naturally smooths  the transition region and provides superior flexibility when fitting intermediate spiral galaxies (e.g., NGC~2403, NGC~2903).
Finally, the  BEC DM profile drops abruptly to zero  density at $R_h$  \cite{Bohmer2007,Chavanis2011}, quantum pressure in the LHY system converts the outer edge into an exponentially decaying regime ($\rho \sim \tilde{r}^{-2} e^{-2\kappa \tilde{r}}$).
This ensures a smooth, nonsingular convergence of the total mass, $M_\infty$ 
while the circular velocity rises linearly in the core, reaches a broad maximum and then declines slowly toward the Keplerian fall-off, which mimics the nearly flat curve over the radial range probed by the data.
Overall, the standard BEC profile improves on NFW ($\langle\chi^2_\nu\rangle=3.75$ versus 5.38).
The present LHY DM model yields the lowest sample average, $\langle \chi^2_{\nu} \rangle=1.20$
(median 0.29; $\chi^2_{\nu}<1$ for 13 galaxies, $1-3$ for 5, and $>3$ for CamB and UGC~08490), across dwarf irregular, LSB and massive spiral galaxies.

\begin{figure}
	\includegraphics[scale=0.5]{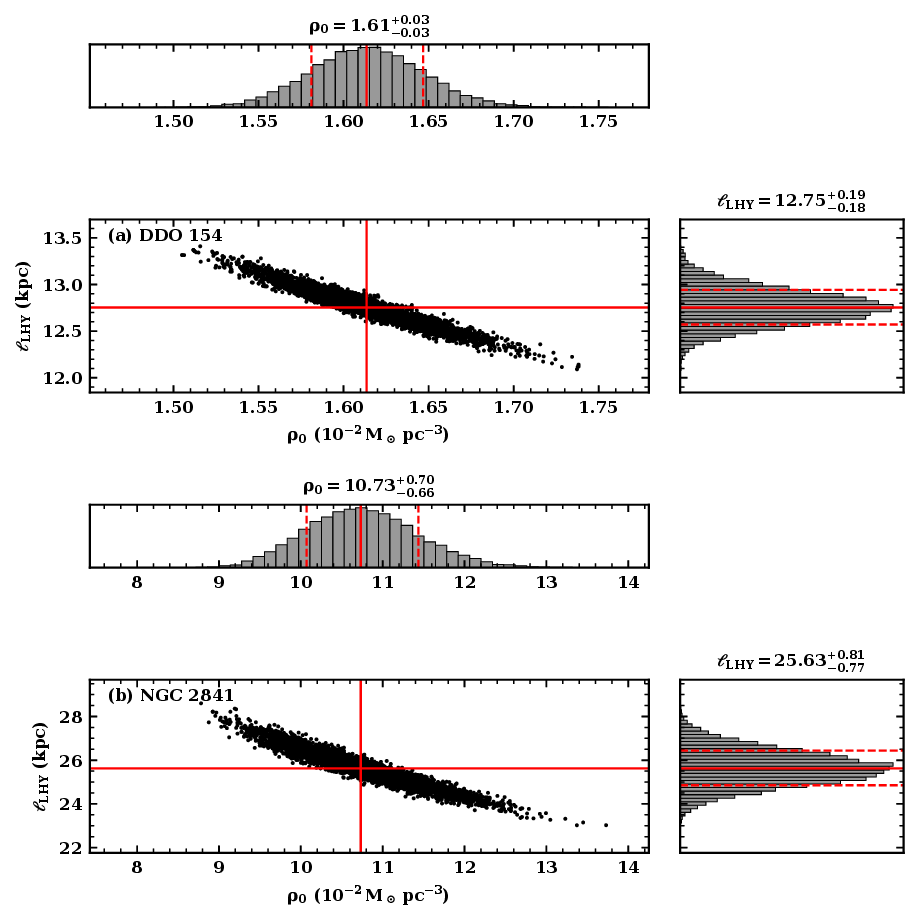}
		\caption{Posterior distributions of the pure-LHY DM halo parameters $(\rho_0, \ell_{\text{LHY}})$  for two representative SPARC galaxies, (a) DDO 154 and (b) NGC 2841, from the ensemble-MCMC chains.  Points are chain sample, solid red lines mark the posterior medians and dashed lines the 16th and 84th percentiles in the marginal histogram.
		The two parameters are anticorrelated. }		
\label{fig4} 
\end{figure}

To examine the parameter space and structural characteristics of the full LHY DM halo, Markov chain Monte Carlo (MCMC) sampling is conducted for all 20 galaxies from the SPARC database. 
The uncertainties and parameter degeneracies are estimated by sampling the posterior of $\theta=(\rho_0,\ell_{\rm LHY})$ with an
affine-invariant ensemble sampler (our implementation of the Goodman-Weare stretch move, i.e. the algorithm of \texttt{emcee}~\cite{ForemanMackey2013}).
We adopt flat priors in $\log_{10}\rho_0\in[-5,2]$ ($M_\odot\,{\rm pc}^{-3}$) and $\log_{10}\ell_{\rm LHY}\in[-1.5,2.5]$ (kpc),
which are also the bounds of the least-squares fits. The likelihood is Gaussian,
$\ln\mathcal L=-\tfrac12\sum_i[(v_{\rm obs}-v_{\rm total})/\sigma_i]^2$, with the quoted $\sigma_i$ and fixed $\Upsilon_{\rm disk},\Upsilon_{\rm bulge}$.
For each galaxy we run 32 walkers for $8000$ steps, started in a small ball around the least-squares optimum; the first
$\max(25\%,5\tau)$ of each chain is discarded. Convergence is judged by the integrated autocorrelation time, requiring a chain
length $>50\tau$, and by the split Gelman--Rubin statistic, requiring $\hat R<1.01$~\cite{Gelman1992}. 
The chains converge for 17 of the 20  galaxies ($\tau\simeq30$-36 steps, $\hat R\le1.008$); CamB, UGC~05918 and UGCA~442 do not (Fig.~\ref{fig4}). 
For the converged galaxies the 1$\sigma$ relative uncertainties are 2.0-11.4\% for $\rho_0$ (mean 6.6\%) and 1.5-10.3\% for $\ell_{\rm LHY}$ (mean 3.8\%).

The obtained joint 2D posterior and 1D marginalized probability distributions for the central mass density $\rho_0$ and quantum scale radius $\ell_{\rm LHY}$ reveal fundamental physical properties of the LHY condensate. 
As shown in Fig.~\ref{fig4}, for both dwarf and spiral systems,  the 2D posterior contours display a distinct negative slope, reflecting an inverse parameter degeneracy ($\rho_0 - \ell_{\text{LHY}}$  anti-correlation). 
Within the inner core region ($r \lesssim \ell_{\rm LHY}$), the  circular velocity scales approximately as  $v_{\text{DM}} \simeq  \sqrt{4\pi G \rho_0/3}\, r$ which fixes $\rho_0$ independently of $\ell_{\rm LHY}$. 
The anticorrelation instead reflects the amplitude of the rotation-curve maximum $v_{\max}\propto\sqrt{\rho_0}\,\ell_{\rm LHY}$: a higher $\rho_0$ can be compensated by a smaller $ \ell_{\rm LHY}$ while keeping $v_{\rm max}$, and hence the outer velocities, unchanged.
Comparing DDO~154 and NGC~2841 highlights a systematic halo expansion across galactic scales. For DDO~154 the sampling yields 
$\ell_{\rm LHY}=12.75^{+0.19}_{-0.18}$~kpc and $\rho_0=(1.61\pm0.03)\times10^{-2}\,M_\odot\,{\rm pc}^{-3}$, whereas NGC~ 2841 requires 
$\ell_{\rm LHY}=25.63^{+0.81}_{-0.77}$~kpc  and higher density  $\rho_0=(10.73^{+0.70}_{-0.66})\times10^{-2}\,M_\odot\,{\rm pc}^{-3}$.

\subsection{Extracted LHY halo parameters}

\begin{table*}[htbp] 
\centering 
\caption{Best-fit LHY DM halo parameters ($\rho_0$ and $\ell_{\rm LHY}$) obtained from fitting the full numerical solutions of the coupled GGPPE against observational rotation curve data for 20 representative SPARC galaxies.
The halo profile is evaluated at $\lambda=4000$  (18 galaxies) or $\lambda=0.05$ (DDO~168, UGCA~442), the limits of the precomputed table, so $\rho_0$ and  $\ell_{\rm LHY}$ are effective scale parameters and do not fix the boson mass.  $R_{\max}$ and  $V_{\rm obs,max}$  refer to the SPARC data
the $\ell_{\rm LHY}$ of CamB lies at  the upper fit bound (316~kpc) and is unconstrained; its error is not meaningful. 
Uncertainties are the least-squares $1\sigma$ errors; MCMC results are in Fig.~\ref{fig4}.}
\begin{tabular}{rllccccccccc}
\hline\hline
\# & Galaxy & Type & $R_{\max}$ & $V_{\rm obs,max}$ & $\rho_0$ & $\ell_{\text{LHY}}$ &  $\chi^2_\nu$ & AIC & BIC \\
 & & & (kpc) & (km s$^{-1}$) & ($10^{-2}\,M_\odot$ pc$^{-3}$) & (kpc) & & & & \\
\hline
1 & CamB & Dwarf Irregular & 1.79 & 20.1 & $0.19\pm0.17$ & $316.23\text{ (bound)}$  & 4.50 & 35.5 & 35.9 \\
2 & DDO 154 & Dwarf Irregular & 5.92 & 48.2 & $1.61\pm0.03$ & $12.75\pm0.18$  & 1.92 & 23.2 & 24.2 \\
3 & DDO 168 & Dwarf Irregular & 4.12 & 55.0 & $2.04\pm0.11$ & $0.30\pm0.01$  & 2.15 & 21.2 & 21.8 \\
4 & IC 2574 & LSB Dwarf & 7.20 & 61.5 & $1.16\pm0.08$ & $14.75\pm0.65$  & 0.04 & 4.5 & 5.9 \\
5 & NGC 2403 & Spiral & 13.37 & 67.5 & $0.40\pm0.02$ & $34.64\pm0.92$ & 0.30 & 12.8 & 15.6 \\
6 & NGC 2841 & Massive Spiral & 15.60 & 312.0 & $10.75\pm0.64$ & $25.62\pm0.75$ & 1.84 & 24.2 & 25.3 \\
7 & NGC 2903 & Spiral & 8.80 & 215.0 & $17.27\pm0.85$ & $15.09\pm0.40$ &  0.31 & 6.8 & 7.6 \\
8 & NGC 3198 & Spiral & 10.50 & 149.0 & $5.64\pm0.26$ & $17.57\pm0.46$  & 2.72 & 36.6 & 37.9 \\
9 & NGC 3521 & Spiral & 11.00 & 225.0 & $16.56\pm1.19$ & $14.95\pm0.50$  & 0.15 & 5.2 & 5.8 \\
10 & NGC 5055 & Spiral & 12.00 & 198.0 & $11.45\pm0.80$ & $15.86\pm0.51$  & 0.04 & 4.4 & 5.0 \\
11 & NGC 6503 & Spiral & 6.00 & 122.0 & $13.91\pm0.80$ & $9.24\pm0.27$ &  1.98 & 19.9 & 20.5 \\
12 & NGC 7331 & Spiral & 14.00 & 261.0 & $15.86\pm1.35$ & $17.61\pm0.67$  & 0.15 & 5.2 & 5.8 \\
13 & UGC 02259 & LSB Spiral & 5.00 & 88.0 & $7.32\pm0.61$ & $8.74\pm0.41$  & 0.21 & 5.7 & 6.3 \\
14 & UGC 02885 & Giant Spiral & 30.00 & 301.0 & $4.77\pm0.54$ & $33.94\pm1.66$ & 0.24 & 6.0 & 6.6 \\
15 & UGC 05918 & Dwarf & 3.07 & 25.0 & $0.98\pm0.22$ & $8.68\pm1.51$ &  0.04 & 4.2 & 3.8 \\
16 & UGC 06614 & LSB Spiral & 15.00 & 189.0 & $4.54\pm0.36$ & $23.88\pm0.96$  & 0.13 & 5.0 & 5.6 \\
17 & UGC 07151 & LSB Spiral & 7.19 & 58.5 & $0.81\pm0.06$ & $25.15\pm2.57$ &  0.76 & 14.6 & 16.2 \\
18 & UGC 08490 & Spiral & 12.33 & 62.2 & $0.68\pm0.04$ & $24.29\pm0.73$ &  6.14 & 40.8 & 41.0 \\
19 & UGC 12810 & High-Mass LSB & 20.00 & 245.0 & $3.92\pm0.30$ & $32.73\pm1.28$  & 0.05 & 4.4 & 5.0 \\
20 & UGCA 442 & Dwarf & 2.98 & 46.9 & $3.81\pm0.69$ & $0.18\pm0.02$ &  0.27 & 7.2 & 8.5 \\
\hline\hline
\label{sparc_lhy_fits}
\end{tabular}
\end{table*}

\begin{table*}[t]
\caption{Goodness-of-fit comparison across 20 representative SPARC galaxies  for
four DM halo models: NFW \cite{Navarro1996,Press2007}, Standard BEC DM \cite{Bohmer2007,Chavanis2011}, QD2M \cite{Boudjemaa2026}, and  the present LHY DM. 
Model selection is quantified using  the AIC ($\mathrm{AIC}=\chi^2_{\min}+2K$), and BIC ($\mathrm{BIC}=\chi^2_{\min}+K\ln N_{\rm obs}$) \cite{Akaike1974, Schwarz1978, Liddle2007},
with $\chi^2_{\min}=\nu\,\chi^2_\nu$, $\nu=N_{\rm obs}-K$, $K=2$ for NFW, standard BEC DM and LHY DM, and $K=3$ for QD2M. Lowest value per galaxy in bold.}
\begin{tabular}{rllcccc@{\hspace{2em}}cccc@{\hspace{2em}}cccc}
\hline\hline
 & & & \multicolumn{4}{c}{$\chi^2_\nu$} & \multicolumn{4}{c}{AIC} & \multicolumn{4}{c}{BIC} \\
\cline{4-7}\cline{8-11}\cline{12-15}
\# & Galaxy & Type & NFW & BEC & QD2M & LHY & NFW & BEC & QD2M & LHY & NFW & BEC & QD2M & LHY \\
\hline
1 & CamB & Dwarf Irregular & 6.03 & 4.50 & 5.25 & 4.50 & $46.2$ & $\mathbf{35.5}$ & $37.5$ & $35.5$ & $46.6$ & $\mathbf{35.9}$ & $38.1$ & $35.9$ \\
2 & DDO 154 & Dwarf Irregular & 11.77 & 11.44 & 27.00 & 1.92 & $121.7$ & $118.4$ & $249.0$ & $\mathbf{23.2}$ & $122.7$ & $119.4$ & $250.5$ & $\mathbf{24.2}$ \\
3 & DDO 168 & Dwarf Irregular & 10.66 & 2.64 & 1.63 & 2.15 & $89.3$ & $25.1$ & $\mathbf{17.4}$ & $21.2$ & $89.9$ & $25.7$ & $\mathbf{18.3}$ & $21.8$ \\
4 & IC 2574 & LSB Dwarf & 3.30 & 0.79 & 1.61 & 0.04 & $46.9$ & $14.3$ & $25.3$ & $\mathbf{4.5}$ & $48.3$ & $15.7$ & $27.4$ & $\mathbf{5.9}$ \\
5 & NGC 2403 & Spiral & 1.15 & 0.72 & 1.60 & 0.30 & $37.3$ & $25.0$ & $50.8$ & $\mathbf{12.8}$ & $40.2$ & $27.8$ & $55.1$ & $\mathbf{15.6}$ \\
6 & NGC 2841 & Massive Spiral & 2.05 & 5.52 & 8.37 & 1.84 & $26.5$ & $64.7$ & $89.7$ & $\mathbf{24.2}$ & $27.6$ & $65.8$ & $91.4$ & $\mathbf{25.3}$ \\
7 & NGC 2903 & Spiral & 9.04 & 4.21 & 7.84 & 0.31 & $85.4$ & $41.9$ & $68.7$ & $\mathbf{6.8}$ & $86.2$ & $42.7$ & $69.9$ & $\mathbf{7.6}$ \\
8 & NGC 3198 & Spiral & 3.86 & 8.48 & 12.63 & 2.72 & $50.4$ & $105.8$ & $145.0$ & $\mathbf{36.6}$ & $51.6$ & $107.1$ & $146.9$ & $\mathbf{37.9}$ \\
9 & NGC 3521 & Spiral & 7.43 & 2.75 & 5.00 & 0.15 & $63.5$ & $26.0$ & $41.0$ & $\mathbf{5.2}$ & $64.1$ & $26.6$ & $41.9$ & $\mathbf{5.8}$ \\
10 & NGC 5055 & Spiral & 11.65 & 1.58 & 3.46 & 0.04 & $97.2$ & $16.6$ & $30.2$ & $\mathbf{4.4}$ & $97.8$ & $17.2$ & $31.1$ & $\mathbf{5.0}$ \\
11 & NGC 6503 & Spiral & 4.90 & 7.94 & 12.36 & 1.98 & $43.2$ & $67.5$ & $92.5$ & $\mathbf{19.9}$ & $43.8$ & $68.1$ & $93.4$ & $\mathbf{20.5}$ \\
12 & NGC 7331 & Spiral & 5.98 & 2.29 & 3.84 & 0.15 & $51.8$ & $22.3$ & $32.9$ & $\mathbf{5.2}$ & $52.4$ & $22.9$ & $33.8$ & $\mathbf{5.8}$ \\
13 & UGC 02259 & LSB Spiral & 2.97 & 1.80 & 3.26 & 0.21 & $27.8$ & $18.4$ & $28.9$ & $\mathbf{5.7}$ & $28.4$ & $19.0$ & $29.8$ & $\mathbf{6.3}$ \\
14 & UGC 02885 & Giant Spiral & 4.21 & 1.35 & 2.25 & 0.24 & $37.7$ & $14.8$ & $21.7$ & $\mathbf{6.0}$ & $38.3$ & $15.4$ & $22.6$ & $\mathbf{6.6}$ \\
15 & UGC 05918 & Dwarf & 0.91 & 0.13 & 0.42 & 0.04 & $7.7$ & $4.5$ & $7.3$ & $\mathbf{4.2}$ & $7.2$ & $4.1$ & $6.6$ & $\mathbf{3.8}$ \\
16 & UGC 06614 & LSB Spiral & 7.38 & 0.73 & 1.83 & 0.13 & $63.0$ & $9.8$ & $18.8$ & $\mathbf{5.0}$ & $63.7$ & $10.4$ & $19.7$ & $\mathbf{5.6}$ \\
17 & UGC 07151 & LSB Spiral & 6.03 & 0.64 & 0.57 & 0.76 & $88.4$ & $\mathbf{13.0}$ & $13.4$ & $14.6$ & $89.9$ & $\mathbf{14.5}$ & $15.7$ & $16.2$ \\
18 & UGC 08490 & Spiral & 2.16 & 15.54 & 24.58 & 6.14 & $\mathbf{17.0}$ & $97.2$ & $128.9$ & $40.8$ & $\mathbf{17.1}$ & $97.4$ & $129.1$ & $41.0$ \\
19 & UGC 12810 & High-Mass LSB & 5.56 & 1.64 & 3.32 & 0.05 & $48.5$ & $17.1$ & $29.2$ & $\mathbf{4.4}$ & $49.1$ & $17.7$ & $30.1$ & $\mathbf{5.0}$ \\
20 & UGCA 442 & Dwarf & 0.55 & 0.28 & 0.30 & 0.27 & $10.6$ & $7.4$ & $9.3$ & $\mathbf{7.2}$ & $11.9$ & $8.7$ & $11.2$ & $\mathbf{8.5}$ \\
\hline
 & Sample average & & 5.38 & 3.75 & 6.36 & 1.20 & 53.0 & 37.3 & 56.9 & 14.4 & 53.8 & 38.1 & 58.1 & 15.2 \\
 & Sum over galaxies & &  &  &  &  & 1060 & 745 & 1138 & 287 & 1077 & 762 & 1163 & 304 \\
 & Galaxies preferred & &  &  &  &  & 1 & 2 & 1 & 16 & 1 & 2 & 1 & 16 \\
 & $K$ & & 2 & 2 & 3 & 2 & 2 & 2 & 3 & 2 & 2 & 2 & 3 & 2 \\
\hline\hline
\label{tab:chi2_comparison}  
\end{tabular}
\end{table*}

As shown in Table~\ref{sparc_lhy_fits}, the central density, $\rho_0$, ranges between
$\rho_0\simeq0.4\times10^{-2}\,M_\odot\,{\rm pc}^{-3}$ (NGC~2403) and $\simeq1.2\times10^{-2}$ for the low-density dwarfs (IC~2574),
and $\rho_0\simeq(11$-$17)\times10^{-2}$ for massive bright spirals such as NGC~2841, 2903, 3521 and 7331. The fitted length
$\ell_{\rm LHY}$ takes two distinct values: $\ell_{\rm LHY}\simeq0.2$-0.3~kpc for DDO~168 and UGCA~442, whose fits sit at the TF edge, and $\ell_{\rm LHY}\simeq9$--35~kpc for the other galaxies, which lie in the large-$\lambda$ regime;
CamB is unconstrained ($\ell_{\rm LHY}$ at the upper bound). The reduced $\chi^2_\nu$ is below unity for 13 of the 20 galaxies and is
largest for UGC~08490 (6.1) and CamB (4.5); values well below one indicate that the quoted velocity errors are generous for these galaxies.
The cored LHY halo therefore reproduces most of the sample, while CamB and UGC~08490 remain poorly described.

A detailed examination of Table~ \ref{tab:chi2_comparison}  reveals key physical distinctions in model performance across the galactic sample. 
For DM-dominated dwarfs (e.g. CamB and DDO~154) the cuspy NFW profile gives poor fits ($\chi^2_\nu=6.0$ and 11.8), because its central
density singularity overestimates the inner rotation-curve slope.
In contrast, the LHY model gives $\chi^2_\nu\leq 2.72$ for 18 of the 20 galaxies (0.04-2.72), as its repulsive quantum-fluctuation pressure ($\propto\rho^{5/2}$)
maintains a flat, non-singular central core ($\mathrm{d}\rho/\mathrm{d}r \to 0$ as $r \to 0$)  that matches the inner linear velocity rise.
Moreover, while the standard BEC DM enforces a sharp outer density truncation by neglecting the spatial quantum pressure gradient ($\nabla^2 \sqrt{\rho}$), 
leading to a higher sample average ($\langle\chi^2_\nu\rangle=3.75$),  the LHY model retains full kinetic terms to generate a smooth exponential tail 
($\rho \sim \tilde{r}^{-2} e^{-2\kappa \tilde{r}}$)  that seamlessly traces extended disk velocities.
Finally, QD2M, which has one more free parameter, does not improve upon standard BEC in this sample ($\langle\chi^2_\nu\rangle=6.36$ versus 3.75),
while the LHY framework achieves the lowest mean ($1.20$), combining a stiff inner core with a smooth, exponentially decaying outer envelope.

\subsection{Microscopic implications and statistical model selection}
\label{subsec:microscopic_implications}

The fitted parameters provide a phenomenological characterization of the effective LHY DM halos and may be used to explore their possible microscopic interpretation. 
In particular, the characteristic length $\ell_{\rm LHY}$ is related to the effective coupling and central number density through  Eq.~(\ref{leg})
$$
\ell_{\rm LHY}=\frac{\hbar}{\sqrt{2m g_{\rm LHY}n_0^{3/2}}},
$$
The sound velocity follows from the compressibility of the adopted EOS,  $P=(3/5)g_{\rm LHY}n^{5/2}$, and is given by
$$
c_s^2=\frac{1}{m}\frac{\partial P}{\partial n}=\frac{3g_{\rm LHY}}{2m}n_0^{3/2}.
$$
Consequently, the characteristic length and the central sound velocity satisfy
$$
\ell_{\rm LHY}=\frac{\sqrt{3}}{2}\frac{\hbar}{m c_s}.
$$
This relation provides a consistent connection between the fitted halo and the effective sound velocity. However, the numerical determination of $c_s$  requires an independent specification of the boson mass.
Thus, the fitted values of $\rho_0$ and $\ell_{\rm LHY}$ do not independently determine the microscopic particle mass.
A microscopic interpretation of  $g_{\rm LHY}$ also requires specifying the underlying bosonic interaction.  For a conventional dilute BEC with repulsive contact interactions, the coupling  constant is  $g=4\pi\hbar^2a/m$, 
where $a$ is the $s$-wave scattering length. The standard LHY correction to the energy is 
$$
\mathcal E_{\rm LHY}=\frac{256\sqrt{\pi}}{15}\frac{\hbar^2}{m} a^{5/2}n^{5/2}.
$$
Matching this expression to the effective interaction energy adopted in the present  model $\mathcal E_{\rm LHY}=(2/5)g_{\rm LHY}n^{5/2}$, gives 
$$
g_{\rm LHY}=\frac{128\sqrt{\pi}}{3}\frac{\hbar^2}{m}a^{5/2}.
$$
This matching is a possible microscopic prescription rather than a derivation of the pure-LHY EOS. In a conventional BEC, the LHY contribution supplements the leading mean-field interaction. Its dominance thus requires an additional physical mechanism that suppresses or compensates the mean-field contribution.   One possible realization for DM requires a dedicated microscopic analysis and is beyond the scope of the present  halo study. \\
Accordingly, the parameters extracted from the SPARC rotation curves should be interpreted as effective halo quantities. They provide a basis for investigating possible microscopic realizations, but should not yet be regarded as direct constraints on the boson mass or scattering length. A consistent microscopic  determination would require specifying the interaction model and testing whether a common set of particle parameters can reproduce the effective coupling inferred across the galaxy sample.

In order to assess the statistical performance of the proposed model, we compare its rotation-curve fits with those obtained using the NFW, standard BEC DM, and QD2M prescriptions. 
As summarized in Table~\ref{tab:chi2_comparison}, the  mean reduced $\chi^2$ values over 20 galaxies are  5.38, 3.75, 6.36, 1.20, respectively. The LHY model therefore provides the lowest  sample-average reduced $\chi^2$  among the four prescriptions considered. This uncovers that the adopted LHY profile can describe the selected rotation curves effectively.

To  rigorously evaluate model performance of our pure LHY DM model, we perform a statistical model selection analysis using the AIC and BIC \cite{Akaike1974, Schwarz1978, Liddle2007}:
\begin{equation}
\text{AIC} = \chi^2 + 2K, \quad \text{BIC} = \chi^2 + K \ln(N_{\text{obs}}).
\end{equation}
The statistical comparison summarized in Table~\ref{tab:chi2_comparison} reveals that the pure-LHY model yields the lowest sample-average  chi-squared value, $\langle\chi_\nu^2\rangle=1.20$ among the four halo prescriptions considered here.
The corresponding averages are 3.75 for the standard BEC model, 5.38 for NFW, and 6.36 for QD2M. Thus, for the selected sample of  20 galaxies, the pure-LHY model provides the closest overall agreement with the observed rotation curves under the adopted fitting procedure and error model. 
The information criteria lead to the same overall ranking. The sample-averaged AIC and BIC values are 14.4 and 15.2 for pure-LHY,  compared with 37.3 and 38.1 for standard BEC, 53.0 and 53.8 for NFW, and 56.9 and 58.1 for QD2M.  So, the LHY has the lowest AIC and BIC in 16 of the 20 galaxies.
These results indicate a clear numerical preference for the pure-LHY prescription within the adopted sample and fitting framework. 
Nevertheless, the information criteria should be interpreted with care because the number of observational points varies between galaxies and the QD2M model contains three fitted parameters, whereas the other models contain two. The comparison therefore relies on consistent per-galaxy likelihoods and parameters counting.  

All in all, the results suggest that the pure-LHY model is a competitive description of the SPARC rotation curves, with the lowest AIC and BIC in the present sample.

\section{Conclusion and outlook} \label{concl}

In this work we proposed a model of DM halos based on a self-gravitating Bose fluid whose effective interaction is governed by LHY quantum fluctuations.
By solving the full coupled GGPPE , we retain the contributions of quantum pressure, self-gravity and the repulsive LHY term within a unified framework.  
This formulation provides a means of examining how beyond mean-field interactions may influence the internal structure and galactic dynamics of quantum DM halos. 

The analytical investigation reveals two important features of the resulting solutions.  Near the center, the density remains finite and its radial derivative vanishes, leading to a cored density profile and a circular velocity that increases approximately 
linearly with  radius.  At large distances, the wavefunction develops an exponentially decaying tail rather than terminating at a sharp boundary.   Consequently, the halo has a smooth outer profile, a finite total mass, and a circular velocity that eventually approaches the Keplerian regime.  These properties distinguish the full quantum model from descriptions in which quantum pressure is neglected and illustrate the importance of  the interplay of quantum pressure, LHY, and  self-gravity in investigating halo structure.

We placed the proposed LHY halo model in the context of three established DM descriptions, namely the NFW, standard  BEC DM and QD2M models. 
These frameworks lead to different predictions for the internal structure and outer boundaries of galactic halos. 
However,  the distinctive feature of our model is  the self-consistent combination of a regular central core with a smooth, spatially extended outer density distribution, without imposing  a sharp halo  boundary.

We have further examined the model's phenomenological implications by comparing its predicted rotation curves with observations for 20 galaxies from the SPARC database.  The fits indicate that the model reproduces a range of observed galactic kinematics, including the approximately linear inner rotation curves associated with a cored density profile.  In the system considered, the fitted profiles also provide a useful comparison with the cuspy NFW framework, notably for galaxies whose inner rotation curves are difficult to reconcile with a steep central density profile.  These results motivate further investigation of the LHY framework as an effective description of galactic DM halos.  
The characteristic halo scales inferred from the fits offer a starting point for exploring possible connections between galactic observations and the microscopic parameters of the underlying quantum fluid. 
Such an interpretation requires particular care, however, because the inferred particle properties depend on the assumed  EOS, the relation between the effective LHY coupling and the microscopic interaction, 
and possible degeneracies among the model parameters. 

In conclusion, self-gravitating quantum Bose fluids stabilized by pure-LHY quantum fluctuations represent  a  consistent alternative to cold collisionless DM with additional structural flexibility. 
Future work will extend  the present LHY model to cosmological structure formation, linear perturbation theory, dynamics, and galaxy cluster lensing profiles to further constrain the microscopic mass and interaction coupling of the underlying ultralight boson particles.




\begin{thebibliography}{99}

\bibitem{Zwicky1933}
F. Zwicky, Helv. Phys. Acta, {\bf 6}, 110 (1933).

\bibitem{Rubin1970}
 V. C. Rubin, and  J. r. W.K. Ford, Astron. J., {\bf 159}, 379 (1970).
  
\bibitem{Bosma1981}
A. Bosma,  Astron. J., {\bf 86}, 1825 (1981).

\bibitem{Sofue2001}
S. Yoshiaki and  V. C. Rubin, Annu. Rev. Astron. Astrophys., {\bf 39}, 137 (2001).

\bibitem{Bertone2018}
  B. Gianfranco and H. Dan, Rev. Mod. Phys., {\bf 90}, 045002, (2018).

\bibitem{Strig} 
L. E. Strigari, Phys. Rept. {\bf 531}, 1 (2013).

\bibitem{Zurek} 
K. M. Zurek, Phys. Rept. {\bf 537}, 91 (2014).

\bibitem{Court} 
S. Courteau et al., Reviews of Modern Physics {\bf 86}, 47 (2014).

\bibitem {Baer}
 H. Baer, K.-Y. Choi, J. E. Kim, and L. Roszkowski, Phys. Rept. {\bf555}, 1 (2015).
 
\bibitem{Aram} 
T. Aramaki {\it et al.}, Phys. Rept. { \bf 618}, 1 (2016).

\bibitem{Arb}
 A. Arbey and F. Mahmoudi, Progress in Particle and Nuclear Physics {\bf 119}, 103865 (2021).
 
\bibitem{Mat} 
T. Matos, L. A. Ureña-López, and J.-W. Lee, Front. Astron. Space Sci. {\bf 11}, 1347518 (2024).

\bibitem{Rubin1980}
V. C. Rubin, W. K. Ford, Jr., and N. Thonnard, The Astrophysical Journal, \textbf{238}, 471--487 (1980).

\bibitem{Clowe2006}
D. Clowe, M. Bradač, A. H. Gonzalez, M. Markevitch, S. W. Randall, C. Jones, and D. Zaritsky, The Astrophysical Journal Letters, \textbf{648}, L109--L113 (2006).

\bibitem{Bertone2005}
G. Bertone, D. Hooper, and J. Silk, Phys. Rept. \textbf{405}, 279--390 (2005).

\bibitem{Planck2020}
Planck Collaboration, N. Aghanim {\it et al.}, Astronomy and  Astrophysics, \textbf{641}, A6 (2020).

\bibitem{Arcadi2018}
G. Arcadi {\it et al.}, European Physical Journal C 78, 203 (2018). Article.

\bibitem{Roszkowski2018}
J. S. Bullock and M. Boylan-Kolchin, Annual Review of Astronomy and Astrophysics {\bf 55}, 343 (2017).

\bibitem{Peed}
P. J. E. Peebles and B. Ratra, Rev. Mod. Phys. {\bf 75}, 559 (2003).

\bibitem{Pad} 
T. Padmanabhan, Phys. Rept. {\bf 380},  235 (2003).

\bibitem{Cop} 
E. J. Copeland, M. Sami and S. Tsujikawa, Int. J. Mod. Phys. D {\bf 15}, 1753 (2006).

\bibitem{Nor}
S. Nojiri and S. D. Odintsov, Phys. Rept. {\bf 505}, 59 (2011).

\bibitem{Wen} 
D. H. Weinberg, M. J. Mortonson, D. J. Eisenstein, C. Hirata, A. G. Riess, and E. Rozo,  Phys. Rept. {\bf 530},  87 (2013).

\bibitem{Noj}
S. Nojiri, S. D. Odintsov, and V. K. Oikonomou, Phys. Rept. {\bf 692}, 1 (2017).



\bibitem{Navarro1996}
J. F. Navarro, C. S. Frenk, and S. D. M. White, The Astrophysical Journal, \textbf{462}(2), 563--575 (1996).

\bibitem{deBlok2010}
W. J. G. de Blok, `\textit{Advances in Astronomy} \textbf{2010}, Article ID 789293 (2010).

\bibitem{Persic1996}
M. Persic, P. Salucci, and F. Stel, Monthly Notices of the Royal Astronomical Society, \textbf{281} 27 (1996).

\bibitem{Hu2000}
W. Hu, R. Barkana, and A. Gruzinov, Phys. Rev.  Lett. \textbf{85}(6), 1158--1161 (2000).

\bibitem{Schive2014}
H.-Y. Schive, M.-H. Liao, T.-P. Woo, S.-K. Wong, T. Chiueh, T. Broadhurst, and W.-Y. P. Hwang, Phys. Rev.  Lett. \textbf{113}(26), 261302 (2014).

\bibitem{Hui2017}
L. Hui, J. P. Ostriker, S. Tremaine, and E. Witten, Phys. Rev.  D.  \textbf{95}(4), 043541 (2017).

\bibitem{Chavanis2025}
P. H. Chavanis, Front. Astron. Space Sci. \textbf{12}, 1538434 (2025).

\bibitem{Bohmer2007}
C. G. B{\"o}hmer and T. Harko, JCAP \textbf{025}, 06 (2007).

\bibitem{Chavanis2011}
P. H. Chavanis, Phys. Rev. D \textbf{84}, 043531 (2011).

\bibitem{SikivieYang2009}
P. Sikivie and Q. Yang, Phys. Rev.  Lett. \textbf{103}, 111301 (2009).

\bibitem{Harko2011}
T. Harko, JCAP \textbf{022}, 1105 (2011).

\bibitem{LHY}
T. D. Lee, K. Huang, and C. N. Yang, Phys. Rev. \textbf{106}, 1135 (1957).

\bibitem{Petrov2015} D. S. Petrov, Phys. Rev. Lett. {\bf 115}, 155302 (2015).

\bibitem{Boudjemaa2026}
A. Boudjem{\^a}a, Phys. Rev. D \textbf{113}, 123031 (2026).

\bibitem{Chavanis2019}
P.-H. Chavanis,  Phy. Rev. D. \textbf{100}, 083022 (2019).

\bibitem{Proukaki6}
N. P. Proukakis,  G. Rigopoulos, and A. Soto, Phys. Rev. D {\bf 108}, 083513  (2023).

\bibitem{Daller}
R. Daller,  Fron. Astron. Space Sc. {\bf 10} ,  1121920  (2023).

\bibitem{HaghaniHarko2026}
Z. Haghani and T. Harko,  Phy. Rev. D. {\bf 113}, 044005 (2026).

\bibitem{Panotopoulos2026LHY}
G. Panotopoulos. Physics, \textbf{8}, 32, (2026).

\bibitem{Jean} J. I. Jeans, Phil. Trans. R. Soc. Lond. A  {\bf 199}, 1 (1902).

\bibitem{Lelli2016}
F. Lelli, S. S. McGaugh, and J. M. Schombert, Astron. J. \textbf{152}, 157 (2016). 
\bibitem{Akaike1974}
H. Akaike, IEEE Transactions on Automatic Control, \textbf{19}, 716 (1974).

\bibitem{Schwarz1978}
G. Schwarz, The Annals of Statistics, \textbf{6}, 461 (1978).

\bibitem{Liddle2007}
Andrew R. Liddle, Mon. Not. R. Astron. Soc. \textbf{377}, 74 (2007).


\bibitem{Griffin} 
A. Griffin and H. Shi, Phys. Rep. {\bf 304}, 1 (1998).

\bibitem {Boudj} 
A. Boudjem\^{a}a, Degenerate Bose Gas at Finite Temperatures, LAP LAMBERT Academic Publishing, New York, (2017).

\bibitem{Boudj1}  
A. Boudjem\^{a}a, Phys. Rev. A {\bf 97}, 033627 (2018).


\bibitem{Press2007}
William H. Press, Saul A. Teukolsky, William T. Vetterling, and Brian P. Flannery, \textit{Numerical Recipes 3rd Edition: The Art of Scientific Computing}, Cambridge University Press, New York, NY, USA, (2007).

\bibitem{Hairer1996}
Hairer Ernst and Wanner Gerhard, \textit{Solving Ordinary Differential Equations II: Stiff and Differential-Algebraic Problems}, Springer Series in Computational Mathematics, Springer-Verlag, Berlin, Heidelberg, (1996).

\bibitem{Ascher1995}
Uri M. Ascher, Robert M. M. Mattheij, and Robert D. Russell, \textit{Numerical Solution of Boundary Value Problems for Ordinary Differential Equations}, Society for Industrial and Applied Mathematics, Philadelphia, PA, USA, (1995).











\bibitem{ForemanMackey2013}
D.~Foreman-Mackey, D.~Hogg, D.~Lang, and J.~Goodman, Publ. Astron. Soc. Pac. \textbf{125}, 306 (2013).

\bibitem{Gelman1992}
A.~Gelman and D.~B. Rubin, Stat. Sci. \textbf{7}, 457 (1992).



\end{thebibliography}
\end{document}